%% file: Review_Paper.tex
\pdfoutput=1

 \documentclass[final,5p,times]{elsarticle}

 \usepackage{graphicx}
 \usepackage{mathrsfs}
 \usepackage{dcolumn}

\usepackage{amssymb}
\usepackage{amsmath}
\usepackage{color}
\usepackage{ulem}

\journal{Physica E}

\begin{document}

\input{Review_Paper_Frontmatter_002}
\tableofcontents 
\input{Review_Paper_Section012_002}

\input{Review_Paper_Section3_002}

\input{Review_Paper_Section456_002}

\section*{Acknowledgements}
This work was supported by the German Science Foundation (DFG
priority program 1285 `Semiconductor Spintronics'), the Federal
Ministry for Education and Research (BMBF NanoQUIT), and Centre
for Quantum Engineering and Space-Time Research in Hannover
(QUEST).
G.M.M. acknowledges support from the Evangelisches Studienwerk.\\

\section*{References}









\bibliographystyle{elsarticle-num}







\end{document}

%% file: Review_Paper_Frontmatter_002.tex
\begin{frontmatter}



\title{Semiconductor Spin Noise Spectroscopy:\\ Fundamentals, Accomplishments, and Challenges}



\cortext[cor3]{Corresponding author}
\author{Georg M. M\"uller}
\author{Michael Oestreich}
\author{Michael R\"omer}
\author{Jens H\"ubner\corref{cor3}}
\ead{jhuebner@nano.uni-hannover.de}

 \address{Institut f\"ur Festk\"orperphysik, Leibniz Universit{\"a}t Hannover, Appelstra\ss{}e 2, D-30167 Hannover, Germany}
\begin{abstract}

Semiconductor spin noise spectroscopy (SNS)  has emerged as a unique experimental tool that utilizes spin fluctuations to provide profound insight into  undisturbed spin
dynamics in doped semiconductors and semiconductor nanostructures. The technique maps  ever present stochastic spin polarization of free and localized carriers at thermal equilibrium via the Faraday effect onto the light polarization of an off-resonant probe laser and was transferred from atom optics to semiconductor physics in 2005. The inimitable advantage of spin noise spectroscopy  to all other probes of semiconductor spin dynamics lies in the fact that in principle no  energy has to be dissipated in the sample, i.e., SNS exclusively yields the intrinsic, undisturbed spin dynamics and promises optical non-demolition spin measurements for prospective solid state based optical spin quantum information devices.  SNS is especially suitable for small electron ensembles as the relative noise increases with decreasing number of electrons. In this review, we first introduce the basic principles of SNS and the difference in spin noise of donor bound and of delocalized conduction band electrons. We continue the introduction by discussing the spectral shape of spin noise and prospects of spin noise as a quantum interface
between light and matter. In the main part, we give a short
overview about spin relaxation in semiconductors and summarize
corresponding experiments employing SNS. Finally, we give in-depth insight into the experimental aspects and discuss possible applications of SNS.
\end{abstract}





\end{frontmatter}

%% file: Review_Paper_Section012_002.tex
\section{Introduction}\label{}
The physical properties of the electron spin in semiconductors have been an active field of research since the pioneering publications of Lampel \cite{lampel:prl:20:491:1968} and Parsons \cite{parsons:prl:23:1152:1969} in the end of the 1960s. The theoretical proposal of the spin field effect transistor by Datta and Das in 1990 \cite{datta:apl:56:665:1990} has boosted the number of research papers in this field, giving rise to the notion of semiconductor spintronics \cite{wolf:science:294:1488:2001,zutic:rmp:76:32:2004,awschalom:natphys:3:153:2007,fabian:aps:57:565:2007}. Some years later, localized electronic spins in semiconductors were suggested as possible qubits for quantum computation \cite{kane:nature:393:133:1998, loss:pra:57:120:1998,imamoglu:prl:83:4204:1999}.

As a matter of fact, investigations of the electron spin dephasing or relaxation times have been an integral part of semiconductor spintronics since the very beginning \cite{lampel:prl:20:491:1968, parsons:prl:23:1152:1969,MeierZakharchenya198411}. However, most experimental probes of semiconductor spin dynamics rely on generation of a non-equilibrium spin polarization \cite{lampel:prl:20:491:1968, parsons:prl:23:1152:1969, heberle:prl:72:3887:1994, baumberg:prl:72:717:1994, worsley:prl:76:3224:1996, kikkawa:prL:80:43131998}. This creation of a spin polarized electron ensemble away from thermal equilibrium is necessarily accompanied by energy transfer to the system which modifies the effectiveness of the different mechanisms of spin dephasing and can---in the worst case---totally obstruct the measurement.
%
%
Fortunately, this fundamental problem can be circumvented according to  the fluctuation-dissipation theorem \cite{einstein:adp:17:549:1905, kubo:repprogphys:29:255:1966} which states that the response of a system  to a perturbation is directly linked to its fluctuations at thermal equilibrium. In other words, the spin fluctuations of an electron ensemble deliver information about the dynamics of the spin system under an infinitesimal  external perturbation which is   not realizable in an experiment \cite{Kos:nature:431:29:2004}.  Such spin noise, i.e., a time-varying stochastic spin polarization, was first predicted by Bloch \cite{bloch:pr:70:460:1946} for a nuclear spin system and first  measured by Sleator \textit{et al.} \cite{sleator:prl:55:1742:1985}. Later, other groups experimentally verified  the fluctuation-dissipation theorem via magnetometric \cite{ocio:jmmm:54:11:1986,  reim:prl:57:905:1986,alba:jap:61:3683:1987} and electrical \cite{israeloff:prl:63:794:1989} measurements of the magnetic noise of spin glasses. More recently, Rugar and co-workers extended the use of spin fluctuations to very small nuclear and electronic spin ensembles by means of magnetic force microscopy \cite{stipe:prl:86:2874:2001,mamin:prl:91:207604:2003,rugar:nat:430:329:2004, budakian:science:307:408:2005,mamin:prb:72:024413:2005}. In 2006, M{\"u}ller and Jerschow utilized nuclear spin noise for magnetic resonance imaging \cite{muller:pnas:103:6790:2006}.

Measurement of spin noise by optical means, i.e, spin noise spectroscopy (SNS), was first carried out in atom optics by Aleksandrov and Zapasskii \cite{aleksandrov:jetp:54:64:1981} in 1981. Spin-orbit coupling gives rise to an  interaction between the electron spin and the photon helicity via the dipole selection rules and thereby maps the spin noise onto the light polarization of an off-resonant    probe laser via the Faraday effect \cite{faraday:philmag:29:153:1846}. Application of off-resonant probe light is an established concept in atom optics to avoid optical pumping and to study optically thick samples \cite{happer:prl:18:577:1967} and several experiments in alkali metal vapors (see, e.g., Refs.~ \cite{kuzmich:pra:60:2346:1999,julsgaard:nat:413:400:2001}) clearly prove that for sufficient detuning from the resonance SNS can be viewed as a quantum non-demolition measurement \cite{Braginsky08011980,braginsky:rmp:68:1:1996,grangier:nature:396:537:1998} of the atomic spin. 

In 2005, Oestreich and co-workers were the first to demonstrate SNS in a semiconductor system, introducing this sensitive and nearly perturbation-free technique to study semiconductor spin physics \cite{oestreich:prl:95:216603:2005}. SNS has been transferred to semiconductor physics rather late since the shorter spin relaxation times in semiconductors compared to atoms require much more sophisticated experimental means \cite{romer:rsi:78:103903:2007}. Since the pioneering work on semiconductor SNS, a considerable number of theoretical \cite{koenig:prb:75:085310:2007, starosielec:apl:93:051116:2008,kos:prb:81:064407:2010} as well as experimental \cite{romer:rsi:78:103903:2007, muller:prl:101:206601:2008,crooker:prb:79:035208:2009,roemer:apl:94:112105:2009,crooker:prl:104:036601:2010,romer:prb:81:075216:2010,mueller:prb:81:121202:2010} papers  have been published demonstrating SNS in three-, two-, and zero-dimensional
semiconductor systems.


\section{Fundamentals of Spin Noise Spectroscopy}\label{sns}
\subsection{Measurement Principle and Experimental Realization}\label{sns:experiment}
\begin{figure}[tb!]
    \centering
        \includegraphics[width=1.00\columnwidth]{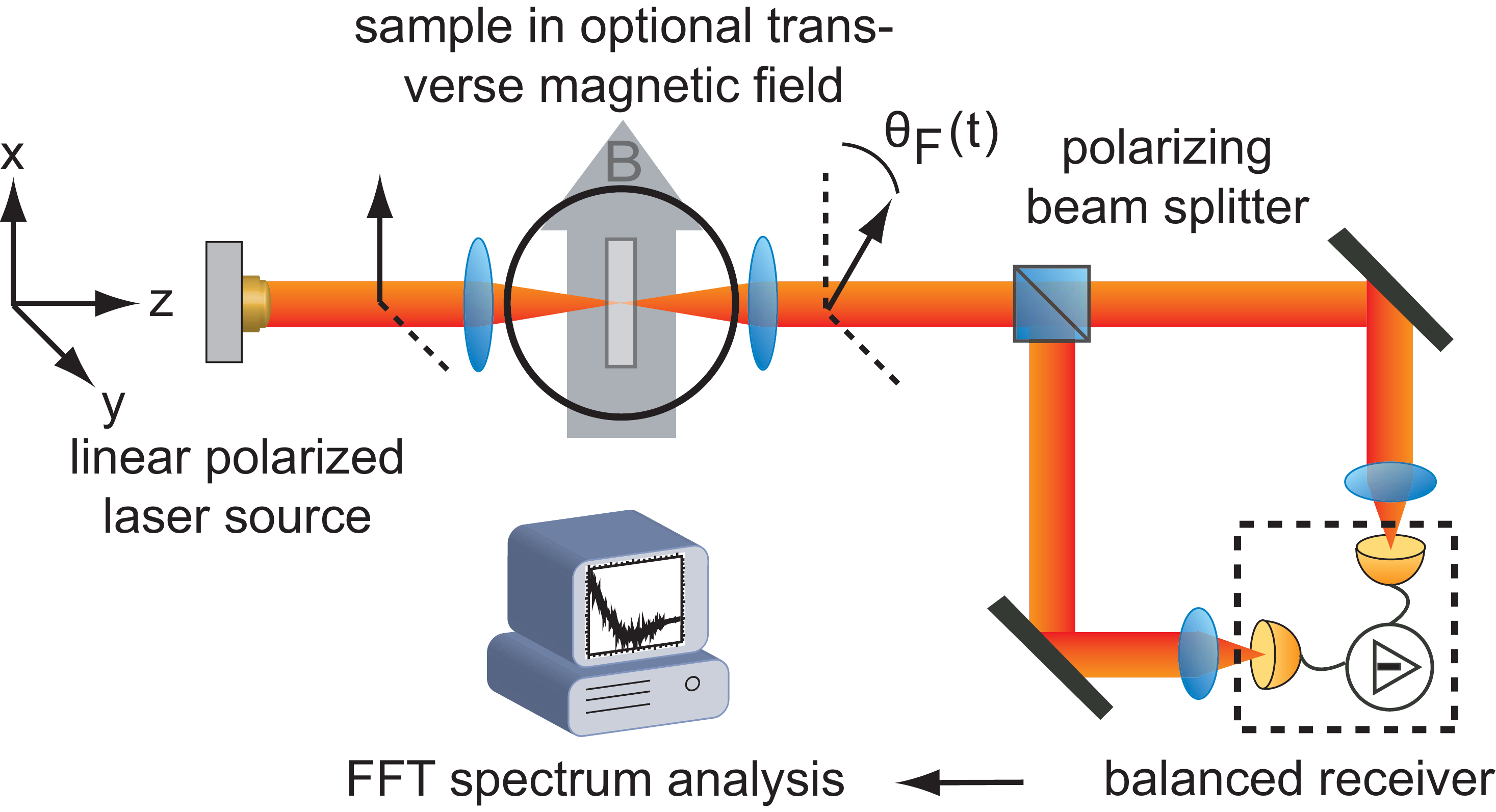}
    \caption{Basic experimental setup for semiconductor SNS. Linearly polarized laser
    light is transmitted through the sample. The Faraday rotation of
    the probe light induced by a stochastic spin polarization is measured via an  optical  polarization bridge. The 
    signal is analyzed in the frequency domain via FFT spectrum analysis.}
    \label{fig:fig4-setup}
\end{figure}
\begin{figure}[t!]
    \centering
        \includegraphics[width=1.00\columnwidth]{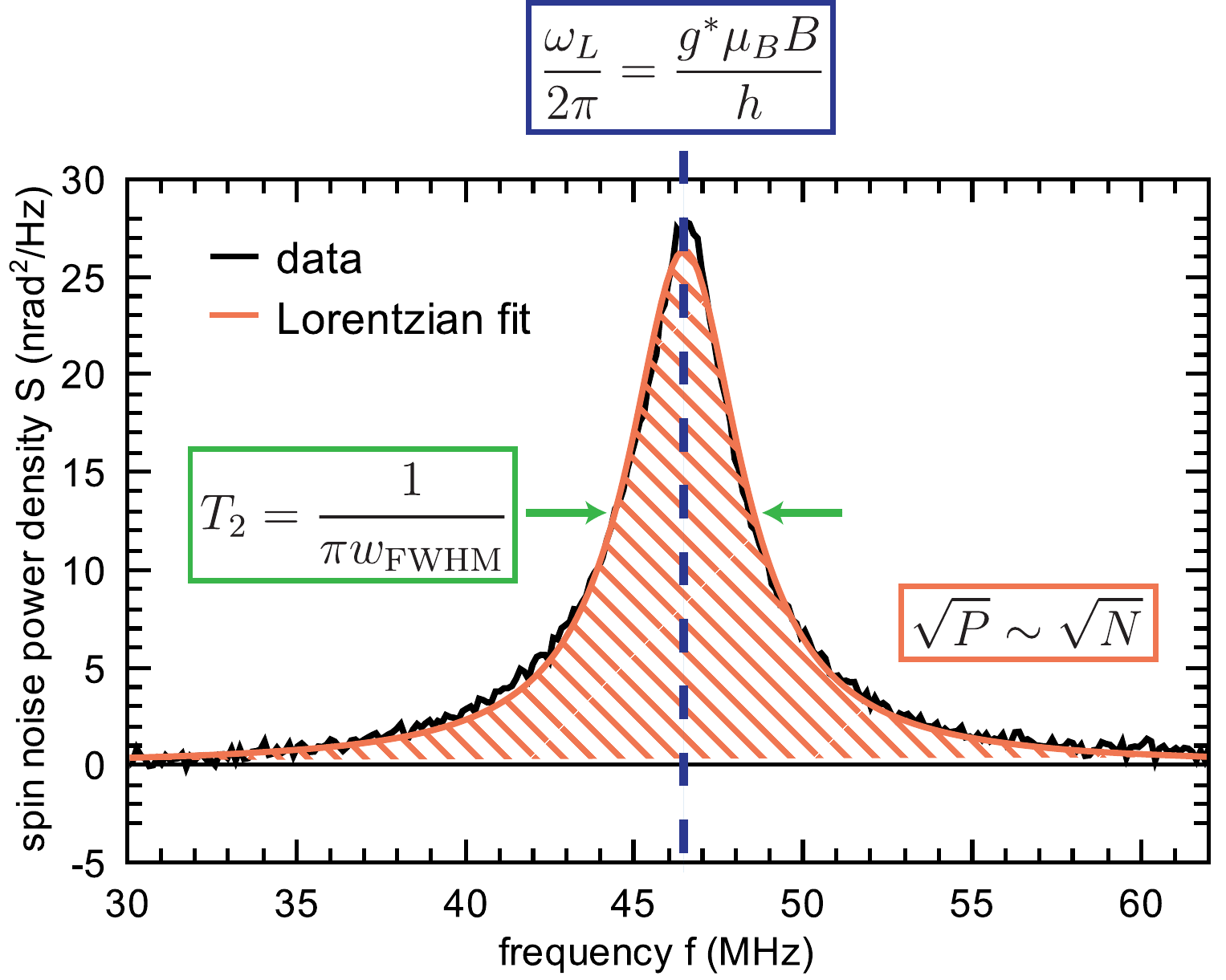}
    \caption{Typical spin noise spectrum with the shot noise background already subtracted. The  
    peak position gives the Larmor frequency and, hence, the effective electron $g$-factor.
    The curve width scales inverse with the spin dephasing time and
    the spin noise power, i.e, the area under the curve, yields valuable
    information about the underlying electron statistics.}
    \label{fig:fig5-lorentz}
\end{figure}

The basic idea of SNS is to map the stochastic spin fluctuations in the sample onto the polarization  of the laser light. Its experimental realization is straightforward and depicted in
Fig.~\ref{fig:fig4-setup}.
Linearly polarized, below band gap laser light is transmitted through the investigated sample, which is mounted in a  cryostat. The carriers in the sample are at thermal equilibrium and, accordingly, the expectation value of the spin polarization $m_z=(N^{\uparrow}- N^{\downarrow})/(N^{\uparrow}+N^{\downarrow})$ vanishes.\footnote{In the following, the axis of quantization for  the electron spin is the $z$-axis, i.e.,
the direction of light propagation (see Fig.~\ref{fig:fig4-setup}). For
the sake of brevity and relevance for the most of the reviewed experiments,
$s=1/2$ electrons are considered. However, SNS is also applicable to hole spins.} Here, $N^{\uparrow}$ and $N^{\downarrow}$ are the number of electrons with spin-up and spin-down, respectively,  and $N=N^{\uparrow}+N^{\downarrow}$ denotes the total number of probed spins. As a finite electron ensemble is considered, the standard deviation $\sigma_{m_z}$  is non-zero. Hence, the electron ensemble shows a  stochastic spin polarization at a given time. This results in a spin dependent bleaching of the probed optical transition and the spin imbalance becomes manifest in a difference in absorption $\alpha_{\pm}$ of right ($\sigma^-$) and left ($\sigma^+$) circularly polarized light which in turn translates via the Kramers-Kronig relations \cite{Kronig:josa:12:547:1926, kramers:acif:2:545:1927} to a
difference of the dispersive part of the refractive indices for the
two circular light components. Due to this circular birefringence, the linearly
polarized probe light, which is composed out of $\sigma^+$ and
$\sigma^-$ light, acquires a rotation of its linear polarization
direction, which is known as the Faraday effect \cite{faraday:philmag:29:153:1846}. Thereby, the spin noise in the sample is projected onto the direction of the linear light polarization. These fluctuations of the Faraday rotation of the probe light are measured in SNS via an optical polarization bridge consisting of a polarizing beam splitter and balanced photodiodes as depicted in Fig.~\ref{fig:fig4-setup}. The  time-continuous electrical signal from the balanced photoreceiver is pre-amplified and---to avoid undersampling (see Sec.~\ref{experiment:averaging})---sent through a frequency bandpass filter  before the analysis in the frequency domain is performed. This spectral analysis reveals the correlations of the underlying spin dynamics. Figure~\ref{fig:fig5-lorentz} shows a typical power spectrum of the measured time  signal with other noise contributions already subtracted (see Sec.~\ref{sns:spurious}). As expounded in detail in Sec.~\ref{sns:spectral}, this spin noise spectrum contains essential information about the spin dynamics: The peak position yields the Larmor frequency $\omega_{L}$, i.e., the precessional frequency of the electronic spins in a transverse magnetic field that is often---but not necessarily---applied in SNS to modulate the spin noise and to shift it from zero frequency. The width of the curve $w_{\text{FWHM}}=\left(\pi T_2\right)^{-1}$ scales with the spin dephasing rate $T_{2}^{-1}$, and the area under the curve gives the spin noise power $P$ which is determined by the number of probed electrons and their degree of  localization. Note that pure homogeneous, i.e., exponential, spin dephasing results in a Lorentzian line shape in the spin noise spectrum whereas inhomogeneous mechanisms yield a Gaussian broadening of the spin noise curve.   To sum up, contrary to conventional experimental probes in which a depolarization or a decay of an artificial spin orientation is measured (see
Sec.~\ref{spindyn:conventional}), in SNS,  no energy has to be deposited in the sample and unperturbed spin dynamics can be experimentally accessed very close at thermal equilibrium.
This measurement principle can be quite universally realized for all kinds of doped semiconductor systems. The probed optical transition is required to be polarizable which means that absorption of circularly polarized light would create a spin orientation. The ratio of the degrees of optical spin orientation and circular light polarization $\xi$ of the investigated transition is determined by the dipole selection rules. In the absence of spin-orbit coupling, every allowed transition would show  vanishing $\xi$ and the resulting spin noise power would be zero. Figure~\ref{fig:fig1-auswahlregeln} schematically depicts the dipole selection rules for the valence to conduction band transition in bulk GaAs. The different optical transition probabilities  of the heavy hole and the light hole to the conduction band with a ratio of $3:1$ yield $\xi =-0.5$ since heavy and light hole bands are degenerate at the $\Gamma$-point and the split-off band is far away in energy and  can therefore be neglected. 

In the following two sections, two complementary systems---a low doped semiconductor with localized, non-interacting electron spins and a highly doped system with delocalized conduction band electrons---are considered. We calculate the mean deviation of $m_z$ and the corresponding spin noise power $P$. We assume that the equilibrium spin polarization due to  magnetic fields is negligible, i.e., $k_{\mathrm{B}}T\gg g^*\mu_{\mathrm{B}}B$ which holds so far for all published semiconductor SNS measurements.

{

\begin{figure}[tb!]
    \centering
        \includegraphics[width=1.00\columnwidth]{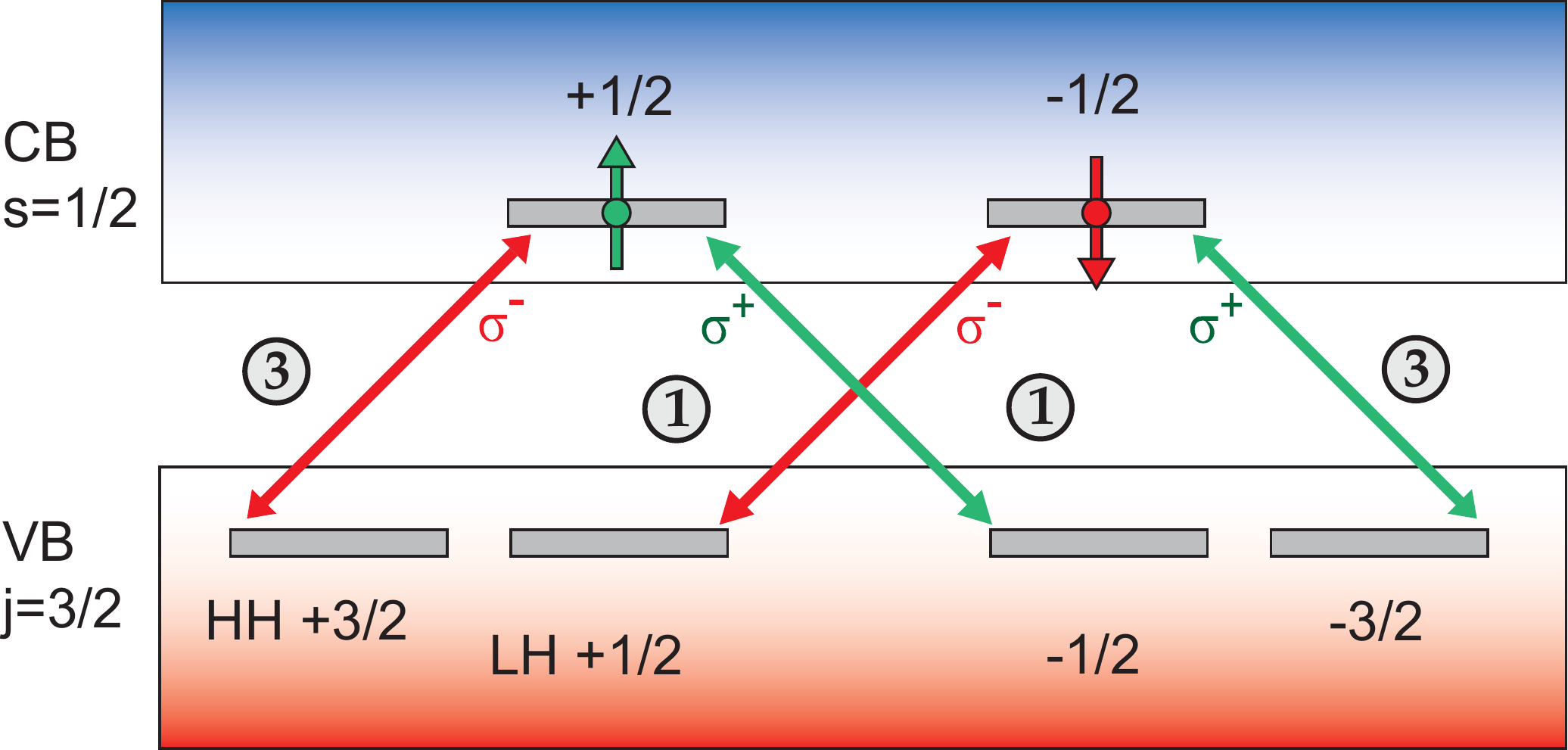}
        \caption{Dipole selection rules for the valence band  (VB) to
        conduction band (CB) transition in bulk GaAs. Left (right) circularly
        polarized light, shown as light (dark) arrows, creates a spin polarization in the
        conduction band of $m_z=-0.5$ ($m_z=+0.5$). The same selection rules also hold for the emission when spin polarized carriers recombine.}
    \label{fig:fig1-auswahlregeln}
\end{figure}
}

\subsection{Spin Noise of Donor Bound Electrons}\label{sns:loc}
In a low doped semiconductor system, where the donor electrons are localized at the impurity atoms and non-interacting with each other, the donor bound exciton transition ($\hbar\omega_{\mathrm{D^0X}}=E_{\mathrm{G}}-E_{\mathrm{D^0X}}^{\mathrm{b}}$) is probed. Here, $E_{\mathrm{G}}$ is the fundamental absorption edge and $E_{\mathrm{D^0X}}^{\mathrm{b}}$ is the binding energy of the exciton-neutral donor complex $\mathrm{D^0X}$ \cite{bogardus:pr:176:993:1968}. Since the donor electrons do not interact with each other, the number of spin-up(down) electrons $N^{\uparrow(\downarrow)}$  follows a binomial distribution with mean $\langle N^{\uparrow(\downarrow)}\rangle = 0.5N$ and standard deviation $\sigma_{N^{\uparrow(\downarrow)}}=0.5\sqrt{N}$. Here, $N=N^{\uparrow}+N^{\downarrow}=n_{\mathrm{D}}V$ is the total
number of donor electrons within the probe volume $V$ with
$n_{\mathrm{D}}$ being the impurity density. Thus, the  standard deviation of the stochastic spin polarization $m_z$ of the donor-bound electrons is given by
\begin{equation}\label{eq:poisson}
\sigma_{m_z}=\sqrt{N}/N.
\end{equation}
 In the absence of spin polarization, the Drude-Lorentz oscillator model describes
this optical transition as a sum of $N$ harmonic oscillators with
oscillator strength $f$. Within this model (see, e.g.,
Ref.~\cite{Demtroder200807}), the   absorption constant $\alpha$ and
the dissipative part of the refractive index $\kappa$ can be written as
\begin{eqnarray}\label{eq:lorentzdissipative}
\alpha& =& \kappa\frac{2\omega}{c_0}\nonumber\\&=&
\frac{e^2fn_{\mathrm{D}}}{8m\epsilon_0\omega_{\mathrm{D^0X}}\sqrt{\epsilon_{\mathrm{B}}}}\frac{\Gamma}{\left(\omega-\omega_{\mathrm{D^0X}}\right)^2+\Gamma^2/4}\frac{2\omega}{c_0},
\end{eqnarray}
where dispersive and dissipative contributions from far-detuned
transitions are subsumed by the background dielectric constant
$\epsilon_{\mathrm{B}}$, and $\Gamma$ is the width of the optical
transition. The corresponding dispersive part of refractive index
is given by
\begin{eqnarray}\label{eq:lorentzdispersive}
n &=& n_{\mathrm{B}}+\tilde{n}\nonumber\\
  &=& \sqrt{\epsilon_{\mathrm{B}}}-
\frac{e^2fn_{\mathrm{D}}}{4m\epsilon_0\omega_{\mathrm{D^0X}}\sqrt{\epsilon_{\mathrm{B}}}}\frac{\omega-\omega_{\mathrm{D^0X}}}{\left(\omega-\omega_{\mathrm{D^0X}}\right)^2+\Gamma^2/4},
\end{eqnarray}
Thus, the deviation of the real part of the refractive index from $n_{\mathrm{B}} = \sqrt{\epsilon_{\mathrm{B}}}$ decreases linearly with inverse
detuning  and the absorbed energy decreases with the inverse
detuning squared. In other words, the change of the refractive index is finite
even at negligible absorption.
\begin{figure}[tb!]
    \centering
        \includegraphics[width=1.00\columnwidth]{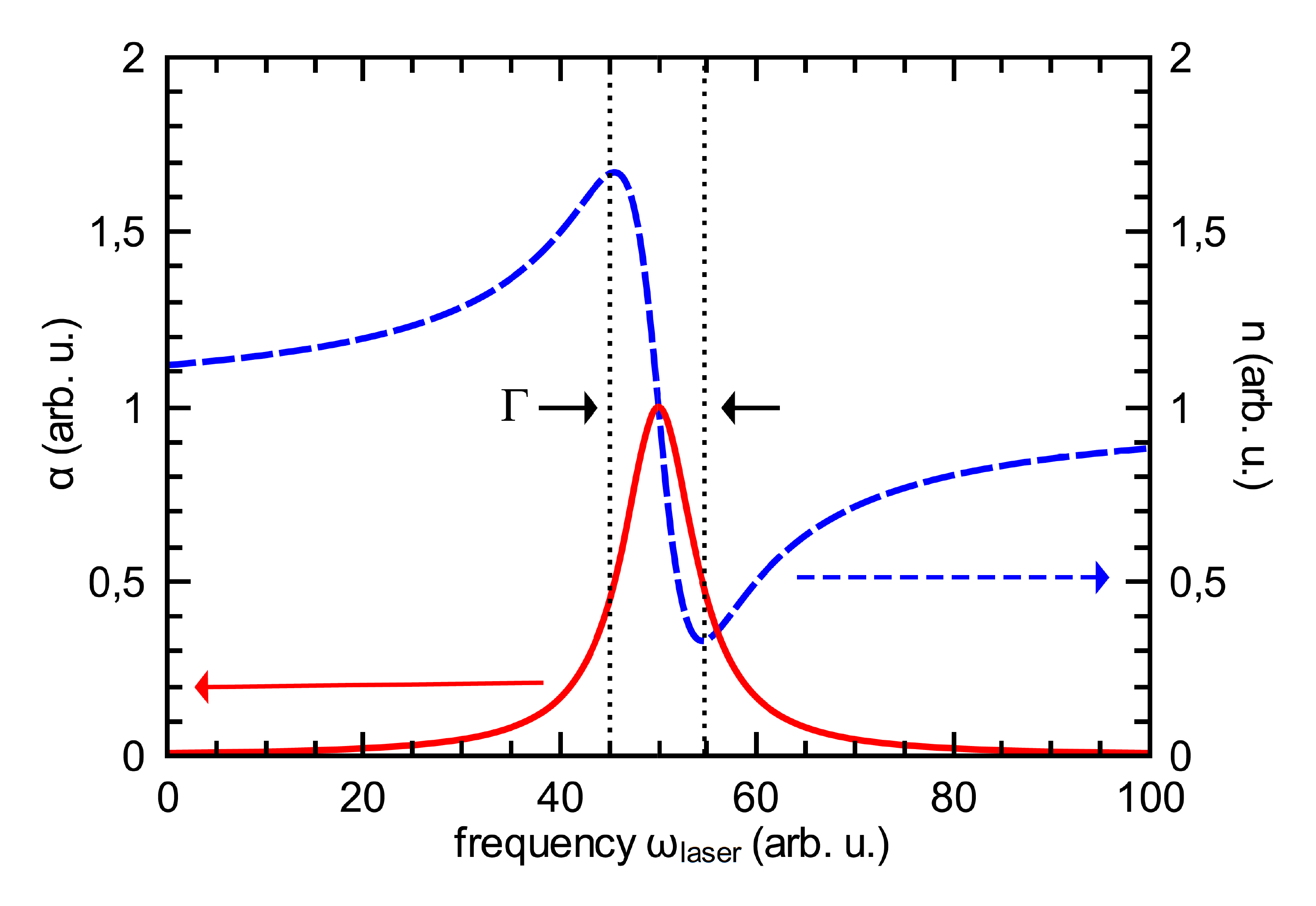}
        \caption{Dissipative and dispersive part of the refractive index
        of an optical transition according to the Drude-Lorentz oscillator
        model. 
       }
    \label{fig:fig2-absorption-refraction}
\end{figure}

The stochastic spin imbalance $m_z$ yields a circular dichroism due to the optical selection rules, i.e., different constants of absorption   $\alpha^+$ and $\alpha^-$  for $\sigma^+$ and $\sigma^-$ light,
\begin{eqnarray}\label{eq:circdichro}
\Delta \alpha &=& \alpha^+-\alpha^- \nonumber\\&=&  \xi \, m_z
\alpha,
\end{eqnarray}
as well as a circular birefringence, i.e., different indices of refraction $n^+$ and $n^-$  for $\sigma^+$ and $\sigma^-$ light,
\begin{eqnarray}\label{eq:circbire}
\Delta n &=& n^+-n^-\nonumber\\ &=& \xi\, m_z\tilde{n}.
\end{eqnarray}
If the  probe beam waist $w_0$ within the sample can be viewed as constant on the
length scale of the sample thickness $l$, the circular birefringence amounts to a
Faraday rotation angle of
\begin{equation}
\theta_{\mathrm{F}}=\pi\,\Delta n \,l / \lambda_0.
\end{equation}
According to Eq.~(\ref{eq:poisson}), the observed integrated spin
noise power reads:
\begin{equation}\label{eq:noisepower}
P=\sigma_{\theta_{\mathrm{F}}}^2=\frac{\pi^2 \xi^2 \tilde{n}^2l}{
\lambda_0^2 n_{\mathrm{D}}A},
\end{equation}
where $A=\pi w_0^2$ is the laser spot area and
$\theta_{\mathrm{F}}$ is assumed to be small. Of course,
Eq.~(\ref{eq:noisepower}) is only valid if the probe volume $V$ is
large compared to the inverse of the doping concentration
$n_{\mathrm{D}}$, which is nearly always the case. 
\begin{figure}[tb!]
    \centering
        \includegraphics[width=1.00\columnwidth]{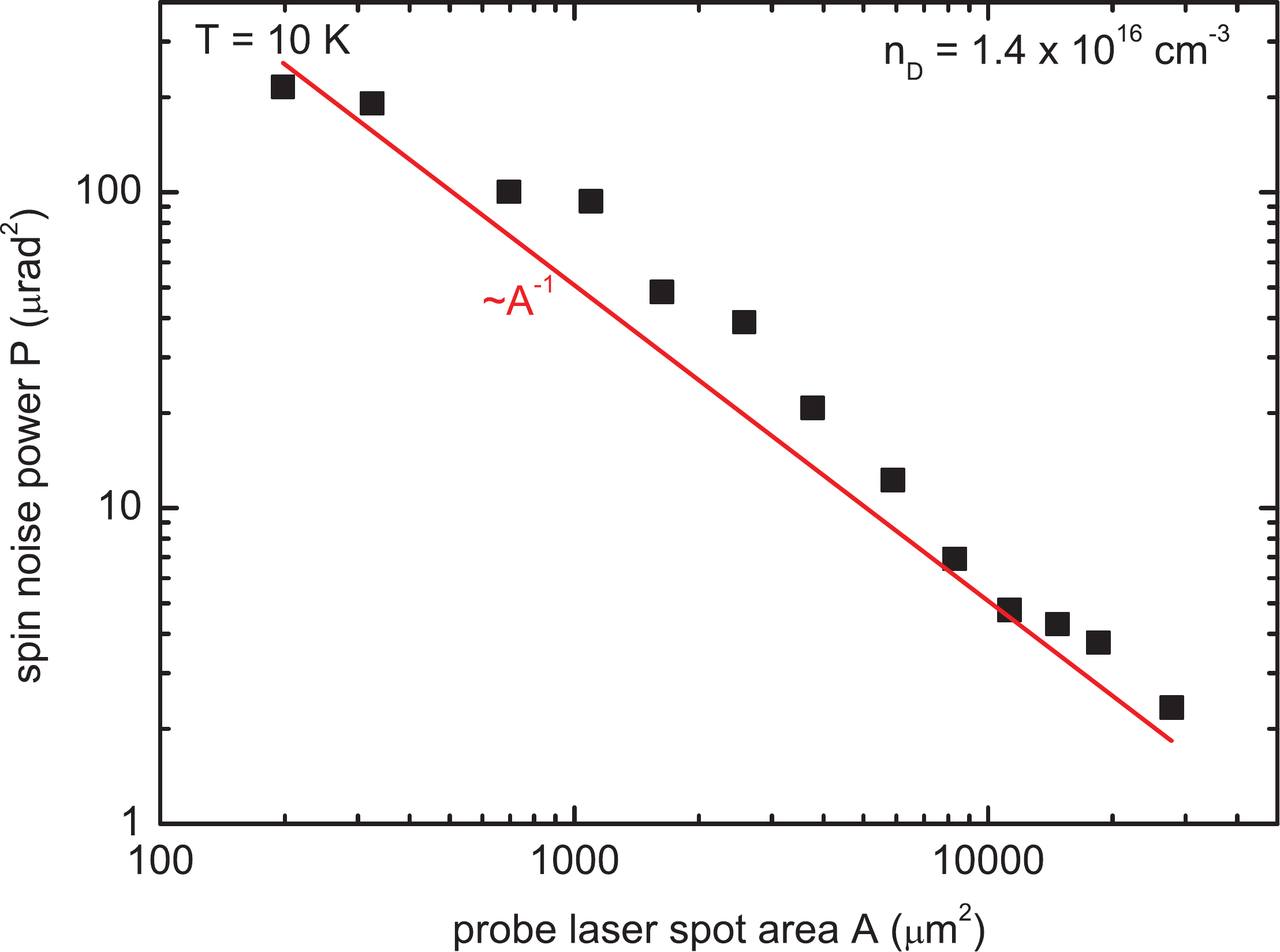}
        \caption{Double logarithmic plot of the integrated spin noise power $P$ as a function of the probe laser spot size $A$ for an $n$-type bulk GaAs sample ($n_{\mathrm{D}}=1.4\times 10^{16}\,\mathrm{cm^{-3}}$) at $10\,$K. The  data is taken from Ref. \cite{crooker:prb:79:035208:2009} and shows the  $1/A$ dependence as expected from Eq. (\ref{eq:noisepower}).
       }
    \label{fig:oneoverarea}
\end{figure}
Note that
Faraday rotation is usually independent of the laser spot area
$A$, but in the case of SNS the  stochastic spin polarization and thereby
the Faraday rotation angle becomes larger for smaller probe volumes, resulting in the $1/A$ relation in the above equation. This behavior of the observed integrated spin noise power was experimentally demonstrated for atomic vapors as well as for semiconductor systems by  Crooker \textit{et
al.} \cite{crooker:nature:431:49:2004,crooker:prb:79:035208:2009} as it is depicted in Fig. \ref{fig:oneoverarea}.

\subsection{Spin Noise of Delocalized Conduction Band Electrons}\label{sns:deloc}
Next, we consider a semiconductor system with a doping level well above the metal-to-insulator transition \cite{mott:rmp:40:677:1968,abram:aip:27:799:1978, shklovskiiefros1984}. In such a sample system, the Fermi energy lies within the conduction band, a behavior that is reported to occur for $n_{\mathrm{D}}^{1/3}a_{\mathrm{B}}>0.43$,  where $a_{\mathrm{B}}$ is the effective Bohr radius of the donor electron-impurity system,  while the critical density of the metal-to-insulator transition  is given by $n_{\mathrm{D}}^{1/3}a_{\mathrm{B}}\approx 0.25...0.33$ \cite{abram:aip:27:799:1978}. The two main differences to very low doped samples are that \textit{(i)} electrons are degenerate and  Fermi-Dirac statistics have to be applied and that \textit{(ii)} the interband transition is described by a sum of Drude-Lorentz oscillators with different resonant energies. This semiconductor system is defined by its density of states $D(E)$ and its doping density $n_{\mathrm{D}}$. The  Fermi level $E_{\mathrm{F}}$ for electrons in the conduction band is calculated via the integral equation
\begin{equation}\label{eq:fermi}
n_{\mathrm{D}}=\int_{E_{G}}^{\infty} f(E)D(E)dE,
\end{equation}
where $f(E)$ is the Fermi-Dirac distribution function
\begin{equation}
f(E)=1/\left(\mathrm{e}^{\left[E-E_{\mathrm{F}}\right]/k_{\mathrm{B}}T}+1\right).
\end{equation}
The stochastic spin polarization in this systems varies
with the energy position in the conduction band. According to the Pauli principle,  the variance of the  spin imbalance is determined by the number of occupied and unoccupied electronic states:
\begin{equation}\label{eq:pauli}
\sigma_{m_z}^2(E)\propto f(E)\left[1-f(E)\right].
\end{equation}
For a more rigorous calculation,  the given  absorption spectrum $\alpha(E)$ has to be modeled by a sum of Drude-Lorentz oscillators with different resonance energies and an energy
dependent spin polarization or even by a fully microscopic model.
\begin{figure}[tb!]
    \centering
        \includegraphics[width=0.6\columnwidth]{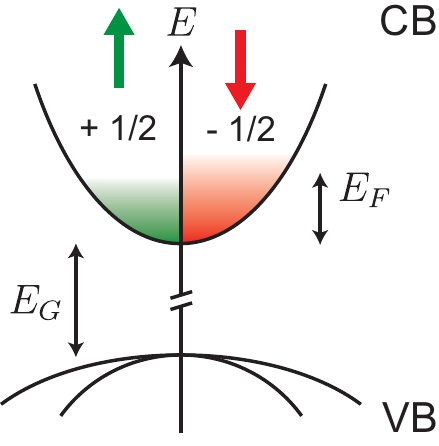}
    \caption{Schematic of the fluctuating occupation numbers  for delocalized spin-up and spin-down electronic states in bulk GaAs.}
    \label{fig:fig3-phasespace}
\end{figure}
From a more practical viewpoint---due to the term $f(E)\left[1-f(E)\right]$---a spin noise contribution is at low temperatures only expected from electrons within a width of $k_{\mathrm{B}}T$ around the Fermi energy $E_{\mathrm{F}}$. Also, the optical absorption sets in at energies around $E_{\mathrm{F}}+E_{\mathrm{G}}$ (see Fig.~\ref{fig:fig3-phasespace}),  again with a width of $k_{\mathrm{B}}T$. Hence, the signal strength can be approximated at low temperatures by calculating the optical transitions by a single Drude-Lorentz oscillator centered around $E_{\mathrm{F}}$  and assuming noise contributions from electrons with a carrier density reduced by a factor $F$ \cite{romer:rsi:78:103903:2007, crooker:prb:79:035208:2009}:
\begin{equation}\label{eq:pauliII}
F\times n_{\mathrm{D}}= \int_{E_{G}}^{\infty} f(E)\left[1-f(E)\right]D(E)dE.
\end{equation} 
The temperature  and probe light wavelength dependence are well described for two- (see Fig. \ref{fig:qw}) \cite{muller:prl:101:206601:2008} as well as for three-dimensional systems \cite{romer:rsi:78:103903:2007} by using this reduced carrier density in Eqs. (\ref{eq:lorentzdispersive}) and (\ref{eq:noisepower}) and modeling the optical transition as described  above.  Besides using the two-dimensional density of states  in Eqs. (\ref{eq:fermi}) and (\ref{eq:pauliII}), also $\xi$ has to be adapted   in order to describe a two-dimensional system, in which the optical selection rules for the three-dimensional case are modified since the degeneracy of heavy and light hole is lifted due to the quantum confinement. Reference \cite{pfalz:prb:71:165305:2005} gives values for $\xi$ for different GaAs/AlGaAs  based quantum well systems.   Note that throughout this section the stochastic spin imbalance is assumed to be small so that the position of the absorption edge is independent of the spin polarization---unlike to magneto-optical measurements in  very high magnetic fields.

At this stage, it is important to note that the amount of observed spin noise power gives information about the underlying electron statistics, i.e., the integrated spin noise power of localized non-interacting electrons is temperature independent while the spin noise power of fully delocalized electrons  vanishes at zero temperature due to Pauli spin blockade \cite{romer:prb:81:075216:2010}. In other words, the amount of spin noise power extrapolated to zero temperature is a measure of the degree of electron localization which is especially interesting for comparative SNS studies in the vicinity of the metal-to-insulator transition (see Sec.~\ref{spindyn:investigation} and Refs.~\cite{crooker:prb:79:035208:2009} and \cite{romer:prb:81:075216:2010}).  

\subsection{Spectral Shape of Spin Noise}\label{sns:spectral}
\begin{figure}[tb!]
    \centering
        \includegraphics[width=1.00\columnwidth]{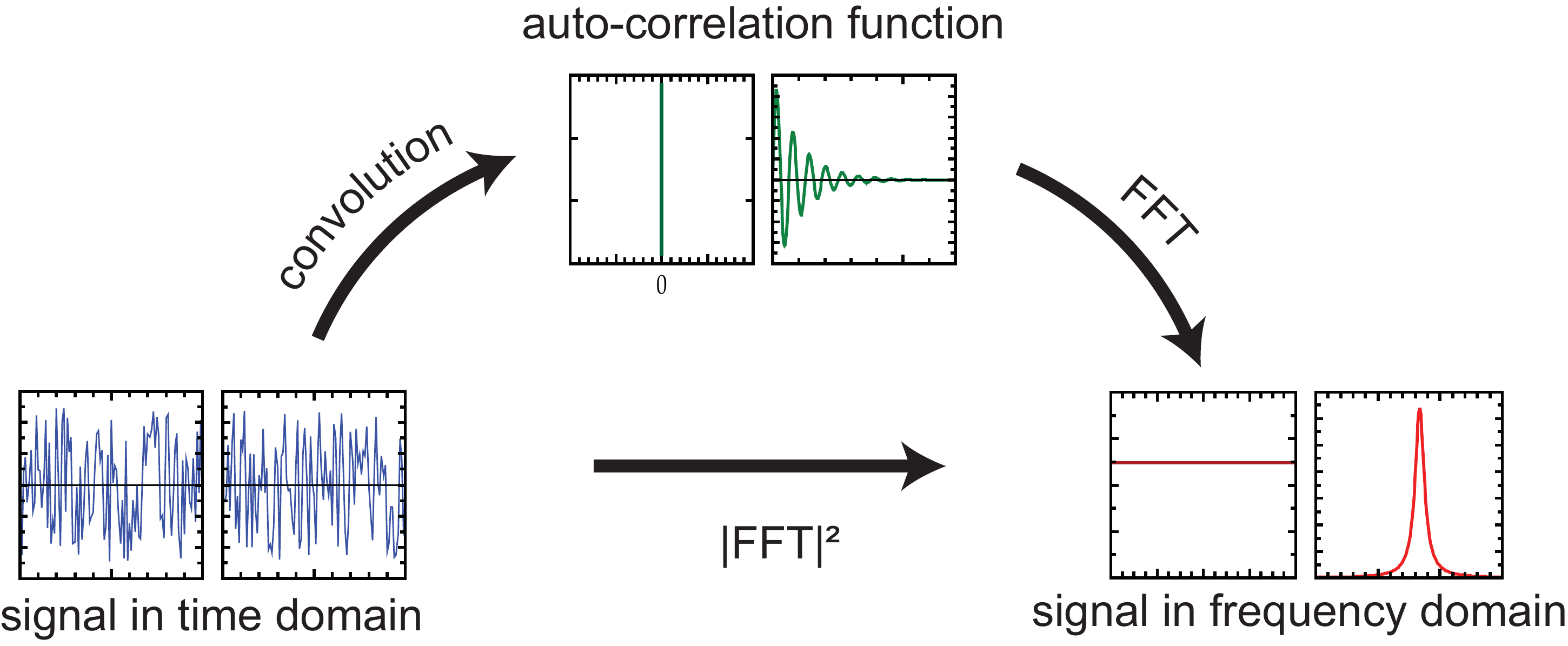}
    \caption{Visualization of the Wiener-Chintchin theorem for  totally uncorrelated noise, like laser shot noise, (left panels) and spin noise (right panels).}
    \label{fig:fig6-fftcircle}
\end{figure}

Figure~\ref{fig:fig5-lorentz} shows a typical spin noise spectrum, i.e., the frequency power spectrum of the spin fluctuations recorded in the time domain. The Wiener-Chintchin theorem  \cite{Wiener1930,Khintchine1934} states that this power spectrum corresponds to the Fourier transform of the auto-correlation function of the time signal.
Sophisticated calculations of the  spin noise spectrum
can be found in Refs.~\cite{koenig:prb:75:085310:2007} and
\cite{kos:prb:81:064407:2010}.
Braun and K\"onig give a fully quantum mechanical density matrix formulation of spin noise   and also consider the special case of an oscillating external magnetic field \cite{koenig:prb:75:085310:2007}; Kos \textit{et al.} \cite{kos:prb:81:064407:2010} explicitly take the orbital motion of the electrons into account in their work and also consider the case of non-negligible transverse spin polarization due to high magnetic fields \cite{kos:prb:81:064407:2010}.\footnote{It directly follows from the spin commutation relations and the Heisenberg uncertainty principle that a  transverse spin polarization increases the uncertainty of the investigated $z$-component (see Sec.~\ref{sns:atomic}). On the other hand, a longitudinal
spin polarization reduces the noise power.} In this paragraph we pursue a more classical approach and consider a single spin precessing in a transverse magnetic field; its auto-correlation function is given by
\cite{koenig:prb:75:085310:2007,kos:prb:81:064407:2010}
\begin{equation}\label{eq:correlation}
\langle s_z(0)s_z(t)\rangle\propto
\cos\omega_{\mathrm{L}}t\,\mathrm{e}^{-t/T_2}\textrm{ for } t>0,
\end{equation}
where $\omega_{\mathrm{L}}$ and $T_2$ are the   Larmor frequency
and the spin dephasing time, respectively.
According to the Wiener-Chintchin
theorem, which is visualized in Fig.~\ref{fig:fig6-fftcircle} for the case of completely uncorrelated,
i.e., white noise (left panels) and spin noise (right panels),
the Fourier transform of this expression directly yields the spin noise spectrum. The Fourier transform of such an exponentially damped spin oscillation yields a Lorentzian shaped
spectral spin noise power density
\begin{equation}\label{eq:powerdensity}
S(f)=\frac{2P}{\pi}\frac{w_{\mathrm{FWHM}}}{4\left(f-\frac{\omega_\mathrm{L}}{2\pi}\right)^2+w_{\mathrm{FWHM}}^2}.
\end{equation}
Here, $w_{\mathrm{FWHM}}=(\pi T_2)^{-1}$ gives the full width at half
maximum (FWHM)  and $P$ is the integrated spin noise power as calculated in Eq.~(\ref{eq:noisepower}).
The experimental data in Fig.~\ref{fig:fig5-lorentz} fits well to a Lorentzian  and, hence,   the Larmor frequency as well as the spin dephasing time can be extracted from the experimental spin noise spectrum.

Nevertheless, the single spin correlation function in Eq.~(\ref{eq:correlation}) is only valid for localized spins. Since the probe volume is always finite,  also spatial correlations have to be considered.  The fact that time of flight broadening modifies the observed spin lifetime in SNS is a known fact from atom optics \cite{katsoprinakis:pra:75:042502:2007}. M\"uller \textit{et al.} were the first to experimentally demonstrate that these transit effects also play an important role in semiconductor SNS \cite{muller:prl:101:206601:2008}. Time of flight broadening is  included in the spin noise spectra by incorporating the spatial degrees of freedom in the spin correlation function $\langle s_z(t_0, \mathbf{r}_0)s_z(t,\mathbf{r}) \rangle$.  This problem was tackled first in a straightforward approach in Ref.~\cite{muller:prl:101:206601:2008} and recently in a more rigorous treatment by Kos \textit{et al.} \cite{kos:prb:81:064407:2010}. However, extracting the spin lifetime and transit time simultaneously from the experimental spin noise spectra is still a very tedious task and more theoretical work is clearly needed here. The essential point is that in many cases experimental access to intrinsic spin lifetimes is only granted by enlarging the probe laser spot \cite{muller:prl:101:206601:2008,crooker:prb:79:035208:2009,romer:prb:81:075216:2010}. In other experimental techniques where oriented spins are continuously optically injected, these effects generally  play a less important role \cite{crooker:prb:79:035208:2009} (see Sec.~\ref{spindyn:conventional}). Nevertheless, these transit effects give the unique possibility to study spatial electron or spin dynamics  at thermal equilibrium in the absence of any gradients in electron or spin density \cite{muller:prl:101:206601:2008}. It should be noted that in principle also pure spatial spin dynamics without accompanying charge dynamics can contribute to this time of flight broadening. Such dynamics were studied by means of spin density gratings, however, necessarily in the presence of spin density gradients \cite{cameron:prl:76:4793:1996,weber:nature:437:1330:2005}.

\subsection{Spurious Noise Contributions}\label{sns:spurious}
The balanced detection  scheme in Fig.~\ref{fig:fig4-setup} is
employed to efficiently reject classical noise of the laser system
due to intensity fluctuations. In addition to this suppressible classical
noise, quantum mechanical shot noise, which is a consequence of the
photon nature of light, contributes to the measured
polarization noise.
In contrast to spin noise, photon shot noise is uncorrelated and therefore results in white noise. Figure~\ref{fig:fig6-fftcircle} illustrates how shot noise becomes manifest in theory as a constant offset in the noise spectrum. In practice, the white shot noise is distorted due to the frequency-dependent sensitivity of the detection system (see Sec.~\ref{experiment:substract} for resulting experimental implications). The optical shot noise level is calculated by Poisson statistics, i.e., the photon flux fluctuates with a standard deviation of $\sqrt{P_{\mathrm{laser}}/\hbar\omega_{\mathrm{laser}}}$ resulting in a  shot noise power density at the detector of $2\nu^2\hbar\omega_{\mathrm{laser}} P_{\mathrm{laser}}$ where $\nu\,\mathrm{[V/W]}$ is the conversion gain of the detector and $P_{\mathrm{laser}}$ and $\omega_{\mathrm{laser}}$ are the probe laser power and energy, respectively.  Checking this linear relationship between the background noise level and the laser power is a quick test that the experimental accuracy is shot noise limited, i.e, at the standard quantum limit. Furthermore, the electronic components, like detector and the amplifier, introduce additional electrical noise. However, in experiments at moderately high laser powers, this noise contribution is usually significantly smaller than the optical shot noise level.

\begin{table*}
 \caption{\label{tab:SNSdata}(a) Magnitude of spin noise power $P$, spin dephasing time $T_2$, peak spin noise power density $S\left(\omega_{\mathrm{L}}/2\pi\right)$,  probe laser power  $P_{\mathrm{laser}}$, and ratio of $S\left(\omega_{\mathrm{L}}/2\pi\right)$ to the background shot noise level, i.e., the signal strength $\eta$, for   different semiconductor systems at cryogenic temperatures. All quantities are detector independent and mostly apply to the weakly perturbing detection regime, i.e., strong detuning and large laser spots (contrary, e.g., to the spectrum in Fig.~\ref{fig:fig5-lorentz}). Electrical noise is neglected for calculating $\eta$. 
(b) Estimated values for prospective measurements. In the case of the single spin, additional electrical noise equivalent to 0.04~mW probe laser power is assumed.}
 \begin{tabular}{lrrrrrr}
 Investigated system &	Ref. & $P$  $\left[\mathrm{\mu rad}^2\right]$ &	$T_2$ $\left[\mathrm{ns}\right]$  & 	$S\left(\omega_{\mathrm{L}}/2\pi\right)$ $ \left[\frac{\mathrm{n rad}^2}{\mathrm{Hz}}\right]$ &$P_{\mathrm{laser}}\,\mathrm{\left[mW\right]}$		& 	$\eta$ \\ \hline
 (a)&&&&&\\
 $n$-doped Bulk GaAs ($1.8\times10^{16}\mathrm{cm^{-3}}$, $l\approx 340\,\mathrm{\mu m}$)&\cite{romer:prb:81:075216:2010}&10&100&1&1&$10^{-2}$\\
  $n$-doped Bulk GaAs ($2.7\times 10^{15}\mathrm{cm^{-3}}$, $l\approx 500\,\mathrm{\mu m}$)&\cite{romer:prb:81:075216:2010}&100&10&1&1&$10^{-2}$\\
 InGaAs quantum dots&  \cite{crooker:prl:104:036601:2010}&1000&1&1&1&$10^{-2}$\\
  $n$-doped Bulk GaAs ($8.8\times 10^{16}\mathrm{cm^{-3}}$, $l\approx 370\,\mathrm{\mu m}$)&\cite{romer:prb:81:075216:2010}&10&10&0.1&1&$10^{-3}$\\
   Modulation $n$-doped multiple quantum well &\cite{muller:prl:101:206601:2008}& 10& 10&0.1&1&$10^{-3}$\\
  $n$-doped Bulk GaAs ($ 10^{14}\mathrm{cm^{-3}}$, $l\approx 2\,\mathrm{\mu m}$)&\cite{romer:prb:81:075216:2010}&$10$&1&$0.01$&1&$10^{-4}$
  
  \\
   \hline
 (b)&&&&&
  \\
Room temperature SNS in bulk GaAs&\cite{oertel:apl:93:132112:2007}&$10$&0.1&$0.001$&1&$10^{-5}$
\\
 Single electron spin in optical cavity&\cite{berezovsky:science:324:1916:2006}&$100$&10&$1$&0.01&$10^{-5}$
 \end{tabular}
 \end{table*}
The  peak spin noise power density at the detector is according to Eq.~(\ref{eq:powerdensity}) given by $S\left(\omega_{\mathrm{L}}/2\pi\right)\times\left(\nu P_{\mathrm{laser}}\right)^2= 2PT_2\times\left(\nu P_{\mathrm{laser}}\right)^2$ where the Faraday rotation is converted into units of the detector output by multiplication with the detector gain $\nu$ and the probe laser power $P_{\mathrm{laser}}$.  Note that the spin noise power is contrary to the shot noise power quadratic in laser power and  a higher laser power accordingly increases the signal strength $\eta$ which is quantified by the ratio between  peak spin noise power density and shot noise power level:
\begin{equation}
\eta=P T_2P_{\mathrm{laser}}/\hbar\omega_{\mathrm{laser}}.
\end{equation}
Table~\ref{tab:SNSdata} gives a survey on recent semiconductor SNS experiments regarding the orders of magnitude of the integrated spin noise power at the detector $P$, the spin dephasing time $T_2$, the peak spin noise power density $S\left(\omega_{\mathrm{L}}/2\pi\right)$, the probe laser power $P_{\mathrm{laser}}$, and the signal strength $\eta$ and also shows estimated values for prospective SNS measurements in a single quantum dot and in bulk GaAs at room temperature. The last column of Tab.~\ref{tab:SNSdata} reveals that efficient data averaging is crucial for SNS in semiconductors, especially for systems with high spin dephasing rates. Therefore, the data acquisition and the subsequent spectral analysis are further discussed in Sec.~\ref{experiment:averaging}.

\subsection{Spin Noise in Atomic Gases: A Quantum Interface between Light and Matter}\label{sns:atomic}
This section gives a brief  and not comprehensive survey on the progress of SNS in atomic vapors of alkali metals that has been achieved since the seminal work of Aleksandrov and Zapasskii \cite{aleksandrov:jetp:54:64:1981}. This discussion mainly aims at the aspect of the quantum non-demolition nature of this experimental method, in which sample excitations are clearly reduced by probing with off-resonant light
\cite{happer:prl:18:577:1967}.
While Aleksandrov and Zapasskii observed Raman like scattering
events in their experiment
\cite{aleksandrov:jetp:54:64:1981,gorbovitskii:optspec:54:229:1983}
indicating light-induced spin flips, several more recent
experiments
\cite{kuzmich:pra:60:2346:1999,kuzmich:prl:85:1594:2000,julsgaard:nat:413:400:2001}
showed that energy absorption is negligible at sufficient detuning
so that these experiments can be viewed as  quantum
non-demolition measurements
\cite{Braginsky08011980,braginsky:rmp:68:1:1996,grangier:nature:396:537:1998}
of the $z$-component of the atomic ensemble spin  $j_z$ with the
$z$-axis being the axis of light propagation. Accordingly, the
Hamiltonian of the interaction between light and matter can be
written as
\begin{equation}\label{eq:interaction}
H\propto j_z\cdot \sigma_z,
\end{equation}
where $\sigma_z$ is the component of the quantum Stokes operator
describing the circular light polarization (see, e.g., Ref.
\cite{kuzmich:pra:60:2346:1999,takahashi:pra:60:4974:1999,ma:pra:79:023830:2009}).
The important feature is that $j_z$ does commute with the
interaction Hamiltonian. For the actual measurement, a meter
variable is needed that does not commute with $H$. Obviously, in
SNS the meter variable is the linear light polarization
$\sigma_{y}$. Solving the Heisenberg equation of motion for the given interaction by taking advantage of the commutation relation
for the components of $\mathbf{j}$ and $\boldsymbol\sigma$
delivers (see, e.g, Ref. \cite{julsgaard:nat:413:400:2001})
\begin{eqnarray}
\sigma_z^{\mathrm{out}}&=&\sigma_z^{\mathrm{in}},\label{eq:QND1}\\
\sigma_y^{\mathrm{out}}&=&\sigma_y^{\mathrm{in}}+\beta \, \sigma_x^{\mathrm{in}}j_z^{\mathrm{in}},\label{eq:faraday}\\
j_z^{\mathrm{out}}&=&j_z^{\mathrm{in}},\label{eq:QND2}\\
j_y^{\mathrm{out}}&=&j_y^{\mathrm{in}}+\beta' \,
j_x^{\mathrm{in}}\sigma_z^{\mathrm{in}}.\label{eq:backaction}
\end{eqnarray}
Here, Eq. (\ref{eq:faraday}) describes the Faraday rotation of the linearly polarized probe light, the quantum non-demolition nature manifests in (\ref{eq:QND1}) and (\ref{eq:QND2}), and the measurement induced back-action on the transverse spin component is given by Eq. (\ref{eq:backaction}). This back-action on the $y$-component of spin system becomes maximal if the $x$-component of the spin  system is fully polarized, i.e., if, according to the Heisenberg uncertainty principle,  the incommensurability of $j_y$ and $j_z$ becomes maximal. Even in the absence of spin fluctuations due to the very long spin lifetimes compared to the measurement time, the measured Faraday rotation is still subject to noise which results from the projective measurement and is known as projection noise \cite{ItanoPRA1993}. A continuous measurement of $j_z$ results in an  increase of the uncertainty of $j_y$: $\left (\delta j_y^{\mathrm{out}}\right)^2=\left (\delta j_y^{\mathrm{in}}\right)^2+\zeta \left (\delta \sigma_z^{\mathrm{in}}\right)^2$ and, subsequently, due to the Heisenberg principle, in a decrease of the uncertainty of $j_z$. Thus, SNS allows measurement of a spin component with accuracy beyond the standard quantum limit as shown theoretically and experimentally by Kuzmich and co-workers \cite{kuzmich:prl:85:1594:2000,kuzmich:epl:42:481:1998}.   In other words, by undergoing  a  SNS measurement,  the coherent spin state evolves into a squeezed spin state \cite{kitagawa:prl:67:1852:1991,wineland:pra:46:R6797:1992} in analogy to the notion of squeezed light \cite{slusher:prl:55:2409:1985}. This concept of spin squeezing was also applied to the atom clock levels of a mesoscopic ensemble of cold caesium atoms resulting in a metrologically relevant noise reduction of 3.4~dB beyond the standard quantum limit \cite{appel:pnas:106:10960:2009}. Julsgaard \textit{et al.} demonstrated that this quantum non-demolition measurement, consecutively carried out on two different alkali vapor samples, yields entanglement of these macroscopic objects \cite{julsgaard:nat:413:400:2001}. Comprehensively, the interaction Hamiltonian in Eq. (\ref{eq:interaction}) represents a so-called quantum interface between matter and light \cite{hammerer:rmp:82:1041:2010, sherson-2006} which, of course can transfer information in both ways. Via this interface, non-classical light states can be used for spin squeezing \cite{kuzmich:prl:79:4782:1997, sorenson:prl:80:3487:1998, hald:prl:83:1319:1999}, the atomic spin ensemble can also be utilized as a quantum memory of light \cite{julsgaard:nature:432:482:2004}, and quantum teleportation between matter and light becomes feasible \cite{sherson:nat:443:557:2006}.

A transfer of these exciting experiments to  semiconductor spin physics is clearly desirable.  Besides possessing very sharp resonances, easy tunable optical density, and a longer spin lifetime, metal atoms in a vapor gas cell  also carry the advantage of an easy optical access from all three spatial directions. Anyhow, in consideration of the experimental results that are reviewed in this section, it is surprising that in theoretical proposals of light and semiconductor spin entanglement \cite{leuenberger:prl:94:107401:2005, leuenberger:prb:73:075312:2006, grond:prb:77:165307:2008, hu:prb:78:085307:2008, potz:prb:77:035310:2008,seigneur:jap:104:014307:2008}, mostly single spins  and not ensemble spins are considered. Additionally, isotropic exchange interaction in an electronic spin ensemble in a semiconductor \cite{kavokin:sst:23:114009:2008} and the resulting correlated spin relaxation  may allow for measurement precision beyond the standard quantum limit for timescales longer than the spin relaxation time \cite{kominis:prl:100:073002:2008}. Application of squeezed light for SNS in semiconductors is discussed in Ref.~\cite{ginossar:prb:77:035307:2008}.


%% file: Review_Paper_Section3_002.tex
\section{Spin Dynamics in Semiconductor Structures}\label{spindyn}
The main purpose of this section is to highlight the insights SNS
has already given to the understanding of electron spin dynamics
in semiconductors as well as to discuss the potential of SNS as an
ultrasensitive probe. 
To this end, we first discuss the most important mechanisms
contributing to the spin decay in semiconductors in Sec.~\ref{spindyn:deph}; this discussion illustrates that spin
orientation as well as electron heating due to above bandgap light
absorption can significantly alter the observed spin dynamics. In
Sec.~\ref{spindyn:conventional}, a survey of the conventional
experimental probes for semiconductor spin dynamics follows and
reveals in which
cases only SNS
grants experimental access to intrinsic spin dynamics. A detailed
overview of SNS experiments in semiconductors closes this section.

\subsection{Spin Dephasing in Semiconductors}\label{spindyn:deph}
There are several publications surveying the physical mechanisms
of spin decay in semiconductors  in great detail
\cite{
zutic:rmp:76:32:2004,MeierZakharchenya198411,kavokin:sst:23:114009:2008,
Dyakonov200809,wu-2010}. On that account, we  only name
the key facts of spin dephasing in semiconductors which play a
role for the understanding of the SNS experiments.

The Heisenberg picture and the spin commutation relation $\left[s_{\mathrm{x}},s_{\mathrm{y}}\right]=i
s_{\mathrm{z}}$ directly disclose the precessional motion
of the spin observable $\mathbf{s}$ in a magnetic field $\mathbf{B}$ where the
Hamilton operator is given by $H\propto
\mathbf{B}\cdot\mathbf{s}$. Extending this treatment to the
polarization of an ensemble of spins $\mathbf{m}$, the  equations
of motion---known as Bloch equations \cite{bloch:pr:70:460:1946}---become for $\mathbf{B}=B\cdot\mathbf{e}_{\mathrm{x}}$ 
\begin{eqnarray}\label{eq:bloch}
\frac{\partial m_x}{\partial t}&=&\frac{\mu_{\mathrm{B}}g^{\ast}}{\hbar}(\mathbf{m}\times\mathbf{B})_x-\frac{m_x-m_x^0}{T_1},\nonumber\\
\frac{\partial m_y}{\partial t}&=&\frac{\mu_{\mathrm{B}}g^{\ast}}{\hbar}(\mathbf{m}\times\mathbf{B})_y-\frac{m_y}{T_2},\nonumber\\
\frac{\partial m_z}{\partial
t}&=&\frac{\mu_{\mathrm{B}}g^{\ast}}{\hbar}(\mathbf{m}\times\mathbf{B})_z-\frac{m_z}{T_2},
\end{eqnarray}
where $g^{\ast}$ is the effective
electron Land\'e $g$-factor which can in consequence of  spin-orbit
coupling  significantly vary from the free electron
$g$-factor.  The quantity $m_x^0$
describes the equilibrium spin polarization along the external
field. Relaxation to equilibrium $m_x^0$ is accompanied by energy
dissipation while the spin polarization transverse to the magnetic
field usually decays with the energy of the spin system being conserved. The spin dephasing time $T_2$ as well as the spin
relaxation time $T_1$ are  
introduced phenomenologically in Eq.~(\ref{eq:bloch}).\footnote{In this review, the definitions of
$T_1$,$T_2$, and $T_2^{\ast}$ according to
Ref.~\cite{zutic:rmp:76:32:2004} are used.}
 SNS is sensitive to  $T_2$ if no magnetic field is applied along the direction of light propagation. This section mainly focuses on the
discussion of spin dephasing  as there has not been an
investigation of the longitudinal spin relaxation time via
semiconductor SNS yet. However,  the
physical difference between spin dephasing and relaxation blurs at the relatively low
magnetic fields used in most SNS experiments,
resulting  in $T_1 \simeq T_2$. Spin dephasing can either be of
homogeneous or of inhomogeneous nature. Inhomogeneous spin
dephasing is for instance observed when an electronic ensemble is probed in
which all electrons experience different magnetic fields or
have different effective $g$-factors.
Inhomogeneous---contrary to homogeneous---spin dephasing is
reversible which means that it could be eliminated in spin echo
experiments; the corresponding inhomogeneous spin dephasing time is denoted by $T_2^{\ast}$.

In order to discuss  the different mechanisms of spin dephasing, we
adopt the random walk formalism of Pines and Slichter
\cite{pines:pr:100:1014:1955}, which  lucidly displays the main
features of the particular mechanisms. Pines and
Slichter consider a spin in interaction with its environment.
This interaction results in an  average change of the spin direction  by
an angle of $\delta \phi$ in the time span  $\tau_{\mathrm{c}}$
where $\tau_{\mathrm{c}}$ is  the correlation time of the given interaction. The
change of the angle  varies its sign with this time constant
 due to scattering events. The mean square of the rotational phase change of the spin after
time $t$ is given by
\begin{equation}
\left\langle \Delta \phi^2 \right\rangle \sim (\delta
\phi)^2t/\tau_{\mathrm{c}}.
\end{equation}
Pines and Slichter define $T_2$ to be the time after which
$\left\langle \Delta \phi^2 \right\rangle$ reaches unity, a
definition that is closely related to the one in Eq.
(\ref{eq:bloch}). In the following, three cases have to be
considered: In the first case \textit{(i)}, the change of the
rotational frequency occurs during the scattering event itself.
The spin dephasing time becomes \cite{fishman:guy:16:820:1977}
\begin{equation}\label{eq:eytype}
1/T_2\sim (\delta \phi)^2/\tau_{\mathrm{c}}.
\end{equation}
 In the second case \textit{(ii)},  the interaction occurs during the whole time
span of $\tau_{\mathrm{c}}$ resulting in a change of the
precession frequency $\omega$ by
$\delta \omega\sim \delta \phi / \tau_c$. The consequent spin dephasing time reads
\begin{equation}\label{eq:motnarr}
1/T_2\sim (\delta \omega)^2\tau_{\mathrm{c}}.
\end{equation}
Thereby, the spin dephasing becomes   less efficient at shorter
correlation times $\tau_{\mathrm{c}}$. This concept of
motional narrowing was first put forward  by Bloembergen
\textit{et al.} to account for the narrow linewidths found in the
nuclear magnetic resonance spectra  of liquids
\cite{bloembergen:pr:73:679:1948}. In the third case \textit{(iii)},  the
correlation time is large compared to $1/\delta \omega$, i.e., the
spin polarization has decayed before the first scattering event
occurs and $\tau_{\mathrm{c}}$ has to be replaced by $T_2$ in Eq.
(\ref{eq:motnarr}), resulting in
\begin{equation}\label{eq:nomotnarr}
1/T_2\sim \delta \omega.
\end{equation}
In general, all mechanisms of homogeneous spin dephasing can be assigned to one of the
three above cases. In the following, we discuss the most relevant processes.
\paragraph{Elliott-Yafet mechanism}
The Elliott-Yafet (EY) mechanism \cite{elliot:pr:96:266:1954,
yafet:ssp:14:1:1963} is based on the fact that  electronic Bloch states are because of spin-orbit
coupling not pure spin-up  or
spin-down states but superpositions of both, e.g.,
$\Psi_{\mathbf{k}n\uparrow}=\left[a_{\mathbf{k}n}(\mathbf{r})|\uparrow\rangle+b_{\mathbf{k}n}(\mathbf{r})|\downarrow
\rangle\right]\mathrm{e}^{\mathrm{i}\mathbf{k\cdot r}}$. This
admixture of the other spin species is small ($|b|\ll 1$).
Nevertheless, scattering into another $\mathbf{k}$-state comes
along with a finite possibility of a spin-flip.  Correspondingly, the EY
mechanism is of the form given by Eq.~(\ref{eq:eytype})
\cite{zutic:rmp:76:32:2004}:
\begin{equation}\label{eq:ey}
1/T_2^{\mathrm{EY}} \sim \langle b^2 \rangle
/\tau_{\mathrm{p}},
\end{equation}
where $\tau_{\mathrm{p}}$ is the momentum scattering time.
Qualitatively, it does not matter which process gives the main
channel for momentum relaxtion. Either scattering due to impurity
atoms \cite{elliot:pr:96:266:1954}, phonons
\cite{yafet:ssp:14:1:1963}, or electron-electron interaction
\cite{boguslawski:ssc:333:389:1980} lead to spin relaxtion via the
EY mechanism. Obviously, all of these underlying scattering
mechanisms obey a strong energy or temperature dependence and,
consequently, optical excitation will alter the efficiency of spin
dephasing. Nevertheless, not only the correlation time is energy
dependent, but also the size of the spin-down admixture $b$ varies with the electronic energy. For III-V semiconductors, Eq.~(\ref{eq:ey})
becomes \cite{chazalviel:prb:11:1555:1975,
fishman:guy:16:820:1977}
\begin{equation}
1/T_2^{\mathrm{EY}}(E_{\mathbf{k}}) \propto E_{\mathbf{k}}^2/
\tau_{\mathrm{p}}(E_{\mathbf{k}}).
\end{equation}
Recently, Jiang and Wu theoretically studied  the relative strength of the EY mechanism to other mechanisms concluding that the EY mechanism is unimportant in most III-V semiconductors at zero magnetic field \cite{jiang:prb:79:125206:2010}.

\paragraph{Dyakonov-Perel mechanism}
In non-centrosymmetric semiconductor structures,  spin-orbit
coupling becomes also manifest in spin-split energy bands, i.e.,
$E_{\mathbf{k}\uparrow}=E_{\mathbf{-k}\downarrow}\neq
E_{\mathbf{k}\downarrow}$. The lack of inversion symmetry can
either result from bulk inversion asymmetry as in III-V
semiconductors (Dresselhaus spin-splitting)
\cite{dresselhaus:pr:100:580:1955,kane:jpcs:1:249:1957}, from
structure inversion asymmetry as in asymmetrically doped quantum
wells (Rashba spin-splitting)
\cite{rashba:spss:2:1109:1960,bychkov:jpc:17:6039:1984}, or from
interface inversion asymmetry (see, e.g., Ref.
\cite{vervoort:prb:56:R12744:1997}).   This spin splitting is
described by an effective, wave vector dependent magnetic field
$\mathbf{\Omega (k)}/g\mu_\mathrm{B}$ ($H=\hbar \mathbf{s}\cdot
\mathbf{\Omega(k)}$). Hence,  spins of electrons in different
$\mathbf{k}$-states precess around different effective magnetic
field vectors and,  subsequently, a spin polarization dephases due
to the so-called Dyakonov-Perel (DP) mechanism
\cite{dyakonov:spss:13:3023:1972}. The correlation time of this
interaction is again given by the momentum scattering rate
including scattering due to impurities, phonons, and
electron-electron interaction. The relevance of electron-electron scattering to the DP mechanism was pointed out by Wu and Ning \cite{wu:epjb:18:373:2000,wu:jpsj:70:2195:2001} as well as by Glazov and Ivchenko \cite{glazov:jetplett:75:403:2002,glazov:jetp:99:1279:2004}.

 For $\tau_{\mathrm{p}}\ll
1/\omega$, which is usually the case, the DP mechanism is of the
type given by  Eq. (\ref{eq:motnarr}):
\begin{equation}
1/T_2^{\mathrm{DP  \ I}}(E_{\mathbf{k}})\sim
\langle\Omega^2\rangle \tau_{\mathrm{p}}(E_{\mathbf{k}}).
\end{equation}
 The Dresselhaus spin-splitting for bulk semiconductors is cubic in $\mathbf{k}$ \cite{dresselhaus:pr:100:580:1955}. Thus, as for the EY mechanism, not only the momentum relaxation time but also the strength of the spin-orbit coupling becomes energy dependent:
\begin{equation}\label{eq:dpgaas}
1/T_2^{\mathrm{DP \ I}}(E_{\mathbf{k}})\propto
E_{\mathbf{k}}^3   \tau_{\mathrm{p}}(E_{\mathbf{k}}).
\end{equation}
The Dresselhaus spin
splitting is modified in quantum wells where $\mathbf{k}$ in the growth
direction is given by momentum
quantization \cite{dyakonov:ssps:20:110:1986}.
 A special case results for $(110)$ grown
quantum wells where the Dresselhaus field has no in-plane component
and, hence, spins aligned along the growth direction do not
dephase due to bulk inversion asymmetry
\cite{dyakonov:ssps:20:110:1986,ohno:prl:83:4196:1999,dohrmann:prl:93:147405:2004}.
 In systems with very low momentum scattering rates as in high mobility quantum wells at ultralow temperatures \cite{brand:prl:89:236601:2002,stich:prl:98:176401:2007}, the DP process is described by Eq.~(\ref{eq:nomotnarr}):
 \begin{equation}
1/{T_2^{\mathrm{DP \ II}}}\sim \langle\Omega\rangle.
\end{equation}

\paragraph{Bir-Aronov-Pikus process}
The interaction between electrons and holes leads  to
electron spin dephasing via momentum scattering and the resulting
EY mechanism. However, Bir \textit{et al.} showed that in presence
of holes electron spin dephasing due to exchange interaction
between electron and holes is much more efficient
\cite{bir:jetp:42:705:1976}. The strength of this Bir-Aronov-Pikus
(BAP) mechanism  depends on the hole density, the electron-hole
overlap, and the fact whether holes are bound or delocalized. The
BAP process shows distinct regimes with different dependencies on
the hole density \cite[Ch.~3]{MeierZakharchenya198411}.
Qualitatively, in all of these regimes, the efficiency of electron
spin dephasing is increasing with hole density. Also, the
temperature dependence does not follow a simple expression and
varies for the different regimes: Nevertheless, for fixed hole
density, the electron-hole overlap increases with decreasing temperatures 
and, accordingly, the BAP mechanism gets more efficient. While it is clear experimental evidence that the BAP
mechanism significantly contributes to spin dephasing in the absence
of other spin dephasing processes
\cite{dohrmann:prl:93:147405:2004}, its relative strength compared
to other mechanisms has become subject of scientific
discussion (see Refs.~\cite{maialle:prb:54:1967:1996} and
\cite{zhou:prb:77:075318:2008}). The  Pauli
blockade strongly suppresses the BAP spin flip mechanism at low
temperatures according to Zhou and Wu  \cite{zhou:prb:77:075318:2008}.

\paragraph{Spin dephasing by hyperfine coupling}
While in natural silicon only roughly $5\%$ of the silicon nuclei
carry a spin angular momentum, in
GaAs all lattice nuclei have a finite spin. In any case, the
electronic spin interacts with the spins of the lattice nuclei
due to the Fermi contact interaction.  
This hyperfine coupling represents an interface between electronic
and nuclear spins,  as  proposed by Overhauser
in 1953 \cite{overhauser:prb:92:411:1953} for the case of metals. In a semiconductor, an electronic spin polarization creates a nuclear spin polarization on the laboratory
timescale which in turn strongly influences the electronic spin
dynamics
\cite{lampel:prl:20:491:1968,dyakonov:jetp:38:177:1974,paget:prb:15:5780:1977}.
 Spin dephasing due to
hyperfine interaction in semiconductors was first theoretically
investigated by Dyakonov and Perel
\cite{dyakonov:jetp:38:177:1974} and later extensively discussed
by Merkulov \textit{et al.} \cite{merkulov:prb:65:205309:2002}.
Depending on the number of magnetic lattice nuclei and the
extension of the donor wavefunction, a localized electronic spin
interacts with a certain number of nuclear spins $N_{\mathrm{L}}$.
Recalling the  prediction of nuclear spin noise by Bloch
\cite{bloch:pr:70:460:1946},  an average stochastic polarization
of $\sqrt{N_{\mathrm{L}}}$ nuclear spins is present at thermal equilibrium.
This hyperfine interaction leads to an electronic spin precession  with an
average frequency $\langle\Omega_{\mathrm{HF}}\rangle$   in the
nuclear magnetic field, the so called Overhauser field. An
expression to calculate this field is given in Ref.
\cite{merkulov:prb:65:205309:2002} (see also Refs.
\cite{romer:prb:81:075216:2010,braun:prl:94:116601:2005}). The
nuclear spins themselves precess in the magnetic field of the
electron, the so-called Knight field, which is a factor of
$\sqrt{N_{\mathrm{L}}}$ smaller than  the Overhauser field. Hence,
in the first step, the nuclear spin polarization can be viewed as
frozen. The correlation time $\tau_{\mathrm{c}}$ of the hyperfine
interaction is determined by the strength of electronic
localization, i.e, by the time an electronic spin resides at a
certain donor site. Spin diffusion via
exchange interaction  occurs orders of magnitude faster than
electronic hopping in the low doping regime  \cite{kavokin:sst:23:114009:2008} so that
$\tau_{\mathrm{c}}\approx\hbar/J$
\cite{dzhioev:prb:66:245204:2002,kavokin:sst:23:114009:2008}   is
given by the exchange integral between remote donor states $J$. In the
intermediate doping regime below the metal-to-insulator
transition, where
$\langle\Omega_{\mathrm{HF}}\rangle\tau_{\mathrm{c}}\ll 1$, a spin
polarization dephases according to Eq. (\ref{eq:motnarr}) with a
rate of \cite{kavokin:sst:23:114009:2008}
\begin{equation}
\label{eq:hfmotional} 1/T_2^{\mathrm{HF \ I}}\sim
\left\langle\Omega_{\mathrm{HF}}^2\right\rangle\tau_{\mathrm{c}}.
\end{equation}
Therefore, spin dephasing based upon hyperfine interaction becomes less efficient with increasing doping concentrations and is completely negligible in
the metallic state.
  In the regime of very low doping and low temperatures, where electrons are considered as non-interacting and strongly localized, no motional narrowing occurs, i.e., $\langle\Omega_{\mathrm{HF}}\rangle\tau_{\mathrm{c}}\gg 1$ and, due to the stochastic nuclear spin polarization, a spin ensemble is subject to inhomogeneous spin dephasing according to Eq. (\ref{eq:nomotnarr}):
\begin{equation}
\label{eq:hfnomotional} 1/{T_2^{\mathrm{HF \ II}}}^{\ast}\sim
\langle\Omega_{\mathrm{HF}}\rangle.
\end{equation}
However, Eqs. (\ref{eq:hfmotional}) and (\ref{eq:hfnomotional})
only describe the decay of the spin components perpendicular to
the Overhauser field at the particular donor sites. In the absence
of an external magnetic field, the angle between electronic and
nuclear magnetic field is conserved during the electronic spin
precession period.  Hence, one third of the spin polarization of a
spin ensemble does not dephase on the timescale of the electronic,
but of the nuclear  precession period as theoretically proposed by
Merkulov \textit{et al.} \cite{merkulov:prb:65:205309:2002} and
experimentally demonstrated by Braun \textit{et al.}
\cite{braun:prl:94:116601:2005}. Due to the spatial variation of
the electronic wavefunction, the Knight field is spatially
inhomogeneous and different nuclei at a given donor site have
different precessional frequencies. Subsequently, the angle
between electronic and nuclear spin is not conserved on the
timescale of the nuclear spin precession. Thus,  the spin component  randomly aligned with
the nuclear field  undergoes spin dephasing with a roughly
estimated rate of
\begin{equation}
\label{eq:hfsingle} 1/{T_2^{\mathrm{HF \ III}}}\sim
\langle\Omega_{\mathrm{HF}}\rangle / \sqrt{N_{\mathrm{L}}}.
\end{equation}

\paragraph{Spin dephasing by anisotropic exchange interaction}  The exchange interaction is mentioned as an origin of motional narrowing of the hyperfine induced spin dephasing in the last paragraph. However, in semiconductors without spatial inversion symmetry, the ex\-change interaction itself is in connection with spin-orbit coupling a source of spin dephasing for localized electronic spins. Due to spin-orbit coupling and a crystalline structure lacking  spatial inversion, the exchange interaction between two spins is not described by a Hamiltonian of the form $\mathbf{s_1}\cdot \mathbf{s_2}$, but by means of a second rank tensor. The antisymmetric part of this tensor gives rise to an anisotropic exchange interaction or the so called  Dzyaloshinskii-Moriya (DM) interaction \cite{dzyaloshinsky:jpcs:4:241:1958, moriya:pr:120:91:1960}. Kavokin was the first to suggest in 2001 that  spin tunneling from one donor site to  another will in average encounter  a finite rotation of  $\gamma$ due to  this anisotropic exchange interaction  \cite{kavokin:prb:64:075305:2001}.  Hence, the DM interaction gives rise to  spin dephasing   of the type of Eq.~(\ref{eq:eytype}):
\begin{equation}
1/T_2^{\mathrm{DM}}\sim \gamma^2 /\tau_{\mathrm{c}},
\end{equation}
where the time between two spin diffusion events
 $\tau_{\mathrm{c}}\approx\hbar/J$ \cite{dzhioev:prb:66:245204:2002,kavokin:sst:23:114009:2008} is, as in the previous paragraph, given by the isotropic part of the exchange interaction $J$.   Electron hopping contributes to spin dephasing analogously \cite{kavokin:sst:23:114009:2008,shklovskii:prb:73:193201:2006}.

\subsection{Conventional Experimental Probes}\label{spindyn:conventional}
SNS was for the first time applied to
investigate spin dynamics in semiconductors in the year 2005
\cite{oestreich:prl:95:216603:2005}. Decades before, quite a
consistent picture on spin dynamics in semiconductors already
existed. Besides on exhaustive theoretical work, this picture had
been mainly based on optical experiments of the steady state
depolarization carried out in the 1960s and 1970s (see Ref.
\cite{MeierZakharchenya198411}) long before time-resolved
measurement techniques relying on (sub) ps laser light pulses were
introduced. 
Investigation of semiconductor spin dynamics  in the time
domain became feasible with the increasing usage  of these new laser light
sources in the early 1990s  (see Ref. \cite{Rulliere200410}). Nevertheless, up to the year 2005, all
optical techniques for investigating spin dynamics in
semiconductors were based on optical orientation of the electron
spins and, hence, move the sample system away from thermal
equilibrium. Besides these optical techniques, also, electron spin
resonance has evolved into a valuable tool to study electron spin
dynamics in semiconductors. The first semiconductor system to be
studied  was $n$-type silicon in the 1950s
\cite{fletcher:prb:94:1392:1954,honig:pr:96:234:1954,feher:prb:114:1245:1959}.
Later, electron spin resonance was transferred to other
semiconductos like InSb
\cite{bemski:prl:4:62:1960,isaacson:pr:169:312:1968}, GaAs
\cite{duncan:physlett:7:23:1963, seck:prb:56:7422:1997}, and InAs
\cite{konopka:pla:26:29:1967}. However, due to dynamic nuclear
effects and low signal strength extracting spin relaxation times
from spin resonance measurements  is often a very difficult task
\cite{isaacson:pr:169:312:1968, seck:prb:56:7422:1997}. Thus,
resonance is often detected by measuring the degree of
depolarization via photoluminescence
\cite{weisbuch:prb:15:816:1977}, electrical transport
\cite{schmidt:cr:263:169:1966}, or  below band gap Kerr
rotation \cite{kennedy:prb:74:161201:2006}. Especially,
electrically detected spin resonance is highly sensitive and
promises even quantum non-demolition measurement of a single spin
\cite{stegner:natphys:2:835:2006,sarovar:prb:78:245302:2008}. In
general, however, electron spin resonance is a depolarization
measurement and, accordingly, the sample is not at thermal equilibrium
during the experiment \cite{feher:prb:114:1245:1959}. Other
experimental probes for spindynamics in semiconductors are also
based on transport and are left out in this section as
they---contrary to the optical techniques---require device
fabrication. A survey of these electrical techniques can be found
in Ref.~\cite{zutic:rmp:76:32:2004}. In the following, we discuss the optical techniques in view of the  different spin dephasing mechanisms (Sec.~\ref{spindyn:deph}) and show that  SNS is the experimental probe of choice for certain sample systems, like, e.g.,
bulk semiconductors with a doping density below the
metal-to-insulator transition.

\paragraph{Optical Measurements of the steady state depolarization}As discussed in Sec.~\ref{sns:experiment}, irradiation of an intrinsic semiconductor with  circularly polarized above band gap light leads  to a spin polarization  in the conduction band  along the $z$-axis, i.e, the direction of light propagation. The maximum degree of polarization $m_z^{\mathrm{max}}\equiv \xi$ is determined by the dipole selection rules (see Fig.~\ref{fig:fig1-auswahlregeln}). A closer look at the corresponding rate equations reveals that the actual degree of spin polarization in undoped semiconductors reads (see Refs. \cite[Ch.~2]{MeierZakharchenya198411} and \cite{zutic:rmp:76:32:2004})
\begin{equation}\label{eq:spinpol}
 m_z=\frac{\xi}{1+\tau/T_2},
 \end{equation}
 where $\tau$ is the electron-hole recombination time. Hence, the steady state electron spin polarization is an indirect measure of the free electron spin dephasing time \cite{ekimov:jetplett:12:198:1970, zakharchenya:jetplett:13:137:1971}. Since the dipole selection rules are not only relevant for light absorption but also for light emission, the steady state electron spin polarization is experimentally accessed by the degree of circular light polarization of the photoluminescence under the assumption that the hole spin is unpolarized due to very effective hole spin dephasing \cite{hilton:prl:89:146601:2002}. The precessional motion of an electron spin polarization in an external transverse magnetic  field [see Eq.~(\ref{eq:bloch})] yields further depolarization of the optically injected spins along the $z$-axis and the value of spin polarization in Eq.~(\ref{eq:spinpol}) becomes \cite[Ch.~2]{MeierZakharchenya198411}
 \begin{equation}\label{eq:hanle}
  m_z(B) =\frac{m_z(0)}{1+\left(\frac{\mu_{\mathrm{B}}g^{\ast}}{\hbar}BT\right)^2},
 \end{equation}
 where the measured spin lifetime $T$ is composed of the actual spin dephasing time and the electron recombination time:
 \begin{equation}\label{eq:hanlelifetime}
 \frac{1}{T}=\frac{1}{T_2}+\frac{1}{\tau}.
 \end{equation}
   This impact of a magnetic field on the polarization state of luminescent light in mercury vapor was discovered by Wood and Ellett \cite{wood:pr:24:243:1924} and explained  by Hanle in 1924 \cite{hanle:zphys:30:93:1924}. In the year 1969, Parsons was the first to measure the quantity $T$  for free electrons in GaSb by means of this Hanle effect \cite{parsons:prl:23:1152:1969}.

 Hanle type measurements can also be carried out to measure the spin dynamics of donor electrons in weakly $n$-doped semiconductors.  Here, the  recombination of bound electrons is usually more efficient than free electron recombination and the spin lifetime is usually longer than the carrier lifetime so that the spin polarization of the optically generated free electrons yields a spin polarization of the  donor electrons  \cite{dyakonov:jetplett:13:144:1971,ekimov:jetplett:13:177:1971}.  In this case, the  electron recombination rate $1/\tau$  in Eq.~(\ref{eq:hanlelifetime}) is substituted by the  rate of replacement of donor electrons  by optically generated spin polarized electrons:
  \begin{equation}\label{eq:hanlelifetimedoped}
 \frac{1}{T}=\frac{1}{T_2}+\frac{G}{n_{\mathrm{D}}},
 \end{equation}
    where $G$ is the excitation rate of free carriers (see Refs. \cite{dzhioev:jetplett:74:182:2001}, \cite{dzhioev:prb:66:245204:2002} and  \cite[Ch.~2]{MeierZakharchenya198411}). Correspondingly, the measured quantity $T$ becomes strongly dependent on the power of the light excitation. According to the above reasoning, $1/T_2$ can be extracted from the linear extrapolation of the power dependence of $1/T$ to vanishing excitation. Nevertheless, this evaluation method requires that $T_2$ is independent of the excitation power---an assumption that is not generally valid since carrier injection  alters the spin dynamics. Especially, at low temperatures, where  polar optical phonons cannot be activated for carrier momentum relaxation (see, e.g., Refs. \cite{lyon:jlum:35:121:1986, shah:ieeeqe:22:1728:1986}),  and at low doping concentrations in the non-degenerate regime, in which---according to equipartition theorem---the electronic energy  scales linear with temperature, the change of the electronic temperature by optical excitation may have a drastic influence on the observed spin lifetime.   Additionally, due to the continuous electron-hole pair generation, the efficiency of the BAP process, the mechanisms based on the DM interaction as well as the hyperfine spin dephasing in the motional narrowing regime is altered because of the presence of free electrons and holes. While the BAP spin dephasing is enhanced because of the increased hole density, the efficiency of the hyperfine and the DM mechanism is reduced due to averaging of the Overhauser field and the anisotropic exchange interaction, respectively. This averaging over several donor atoms occurs via exchange interaction mediated  spin diffusion. Exploiting this effect by flooding the semiconductor with free electrons, Dzhioev \textit{et al.} demonstrated that the hyperfine interaction induced spin dephasing can basically be switched off \cite{dzhioev:jetplett:74:182:2001,dzhioev:prl:88:256801:2002}. Additionally, Paget showed in 1981 that the presence of optically created electrons also significantly alters the observed spin dynamics via exchange averaging between the localized and free electronic states \cite{paget:prb:24:3776:1981}. Again, the overall influence of free electrons  is largest for low-doped samples at low temperatures. In this carrier regime, short spin dephasing times due to hyperfine interaction may also require excitation densities that are comparable to the equilibrium carrier density to achieve a sufficiently high spin polarization. At these intensities, also spin diffusion may modify the depolarization curves and yield a Hanle width that is---contrary to Eq.~(\ref{eq:hanlelifetimedoped})---quadratic in $G$ \cite{dyakonov:sss:10:208:1976}.    To sum up, because of the temperature and free carrier density dependence of the various spin dephasing mechanisms and the non-equilibrium spin polarization, the linear extrapolation of Eq.~(\ref{eq:hanlelifetimedoped}) to zero excitation density may in many cases not be justified and the equilibrium spin lifetime is not accessible in Hanle-type experiments.

Additionally, Hanle-type measurements in contrast to SNS deliver  no independent
information about the effective $g$-factor and the time constant $T$ [see Eq.~(\ref{eq:hanle})], i.e, one of these two quantities has to be known to determine the other. In III-V bulk semiconductors,
the effective electron Land\'e factor $g^{\ast}$ is known to
exhibit an energy dependence (see
Ref.~\cite{yang:prb:47:6807:1993} and references therein) and,
hence, a doping level dependence as well as a temperature
dependence (undoped GaAs:
Refs.~\cite{oestreich:prl:74:2315:1995,oestreich:prb:53:7911:1996,litivenko:prb:77:033204:2008,zawadzki:prb:78:245203:2008,
hubner:prb:79:193307:2009}, $n$-type GaAs:
Ref.~\cite{mueller:prb:81:121202:2010}). Therefore, precision measurements
of the spin lifetime via the Hanle effect require knowledge about
the $g$-factor which has to be gathered by another experiment.
However, if the recombination and spin dephasing rates are known
or negligible,  Hanle measurements allow to determine
the effective electron Land\'e $g$-factor as demonstrated by
Snelling \textit{et al.} with the first study of $g^{\ast}$ in
GaAs/Al$_x$Ga$_{1-x}$As quantum wells in dependence of the quantum
well thickness \cite{snelling:prb:44:11345:1991} (see Ref.~\cite{yugova:prb:75:245302:2007} for a  more recent study). Furthermore, while Hanle type measurements in
dependence on a longitudinal magnetic field (see
Refs.~\cite[Ch.~3]{MeierZakharchenya198411} and
\cite{dzhioev:prb:66:245204:2002})  yield important insight regarding the
correlation time $\tau_\mathrm{c}$ (see Sec.~\ref{spindyn:deph})
of the underlying spin dephasing mechanism, these depolarization
experiments do not allow to study the influence of a transverse
magnetic field on the spin dynamics.

 The degree of polarization of the
luminescent light has not necessarily to be examined in Hanle-type measurements. For
instance, the depolarization of the optically created spin
orientation can also be measured by means of below band gap
Faraday rotation of an additional probe beam as carried out by
Crooker \textit{et al.} to directly compare spin dephasing times
measured by SNS and Hanle measurements
\cite{crooker:prb:79:035208:2009}. However, the different
influence of spin and electron diffusion in both experiments (see
Sec.~\ref{sns:spectral}) makes the experimental data hard to
compare.

\paragraph{Time-resolved optical measurements} The spin dephasing time can be accessed more directly in a time and polarization-re\-solved  measurement of photoluminescence by means of a pulsed circularly polarized  pumping laser and a streak camera system as first carried out in 1994 \cite{heberle:prl:72:3887:1994}. This technique has several advantages over continuous-wave Hanle type measurements since the effective electron $g$-factor, the recombination rate, and the spin dephasing rate are independently of each other extracted from the experiment.   Furthermore, measurements in zero magnetic field as well as studies of the transverse magnetic field dependence are feasible. Time-resolved Kerr \cite{zheludev:89:823:1994,worsley:prl:76:3224:1996} and Faraday rotation \cite{baumberg:prl:72:717:1994,baumberg:prb:50:7689:1994} techniques, both also introduced in 1994, sample the  birefringence, which results from the initial spin orientation by the pump pulse, via the polarization rotation of a  time-delayed, transmitted (Faraday) or reflected (Kerr) probe pulse \cite{oestreich:prl:75:2554:1995}. While  photoluminescence directly reveals the energy position of the carriers whose spin dynamics are probed, time-resolved Kerr and Faraday rotation data usually needs more interpretation \cite[Chap.~2]{Dyakonov200809}. In general, electron relaxation dynamics are accessible in these experiments by means of the transient change of the reflectivity and the transmission, respectively. It was shown in several publications that interpretation of the time-resolved Kerr rotation data is in many cases not possible without also studying the dynamics of electron relaxation \cite{malinowski:prb:62:12034:2000,eldridge:prb:81:033302:2010}.

Again, as carrier relaxation proceeds usually faster than recombination, these three methods  work for  doped as well as for undoped semiconductor systems. The initial optical spin orientation comes along with all the disadvantages that are discussed in the previous paragraph for Hanle-type measurements.  In time-resolved photoluminescence experiments,  free holes and electrons are---like in Hanle measurements---present in the sample during the whole data acquisition time resulting in spin dephasing due to the BAP process  \cite{dohrmann:prl:93:147405:2004}. Recently, Krau\ss{} \textit{et al.} demonstrated the strong influence of various electronic scattering and screening mechanisms  that determine the observed spin dynamics in time-resolved experiments at elevated excitation conditions \cite{krauss:prb:81:035213:2010}. In doped samples with spin lifetimes much longer than the recombination time, time-resolved Kerr and Faraday rotation techniques could in principle  allow  measurements of the spin dynamics  after the electronic system has equilibrated, as it is realized, e.g., in Refs. \cite{kikkawa:prL:80:43131998,yugova:prl:102:167402:2009,korn:njp:12:043003:2010}. In these experiments, however, the repetition rate of the laser system significantly exceeds the spin dephasing rate so that spin dephasing times are extracted via resonant spin amplification, an extension of the time-resolved measurement principle introduced by Kikkawa and Awschalom in 1998 \cite{kikkawa:prL:80:43131998}, in which a transverse magnetic field is swept while the time delay between pump and probe pulse is kept constant.  During some fraction of the data acquisition time, free carriers are still present due to optical pumping and modify the observed dynamics such that the optical excitation has to be included for explaining the experimental outcome  \cite{putikka:prb:70:113201:2004}.
  Also, the rapid optical excitation leads to carrier heating, as in the case of Hanle measurements. Especially, a high fluence of the pump pulse can lead to generation of a large number of non-equilibrium longitudinal optical phonons which further hinders carrier cooling \cite{zhou:prb:46:16148:1992}.  Therefore, due to ineffective cooling of the electron system by phonons at low temperatures \cite{lyon:jlum:35:121:1986,shah:ieeeqe:22:1728:1986}, ultralow electron temperatures are generally not accessible in time-resolved measurements (see Ref.~\cite{hubner:prb:79:193307:2009} where  the temperature of the measured effective $g$-factor levels off at low temperatures indicating insufficient cooling power of the electron system). Possible pitfalls of these experiments are further listed in Ref. \cite{kavokin:sst:23:114009:2008}. The initial optical orientation can also modify the observed spin dynamics due to enthralling effects resulting from the Hartree-Fock contribution to the electron-electron interaction \cite{stich:prl:98:176401:2007}.

\input{Review_Paper_Section3.3ff_002}

%% file: Review_Paper_Section3.3ff_002.tex
\subsection{Investigations by SNS}\label{spindyn:investigation}
Since 2005 SNS has been used to study spin dynamics in bulk semiconductors \cite{oestreich:prl:95:216603:2005,romer:rsi:78:103903:2007, crooker:prb:79:035208:2009, romer:prb:81:075216:2010, mueller:prb:81:121202:2010} as well as in  two \cite{muller:prl:101:206601:2008} and zero dimensional \cite{crooker:prl:104:036601:2010} semiconductor systems.  SNS does not rely on artificial spin orientation, which can---as discussed in previous section---conceal the equilibrium spin dynamics.   At present, all publications on SNS in semiconductors are focused on III-V-based  samples, which can be viewed as quintessential systems for spintronic research. SNS should, however, be applicable to a large group of different semiconductor systems, direct as well as indirect semiconductors, provided that the probed transition obeys appropriate selection rules (see Sec. \ref{sns:experiment}). In the following paragraphs,  we give a survey on the existing investigations via SNS on $n$-type bulk GaAs, modulation-doped (110) GaAs/AlGaAs quantum wells and unintentionally $p$-doped self assembled
(In,Ga)As/GaAs quantum dots.
\paragraph{Bulk GaAs}
\begin{figure}[tb!]
    \centering
        \includegraphics[width=1.00\columnwidth]{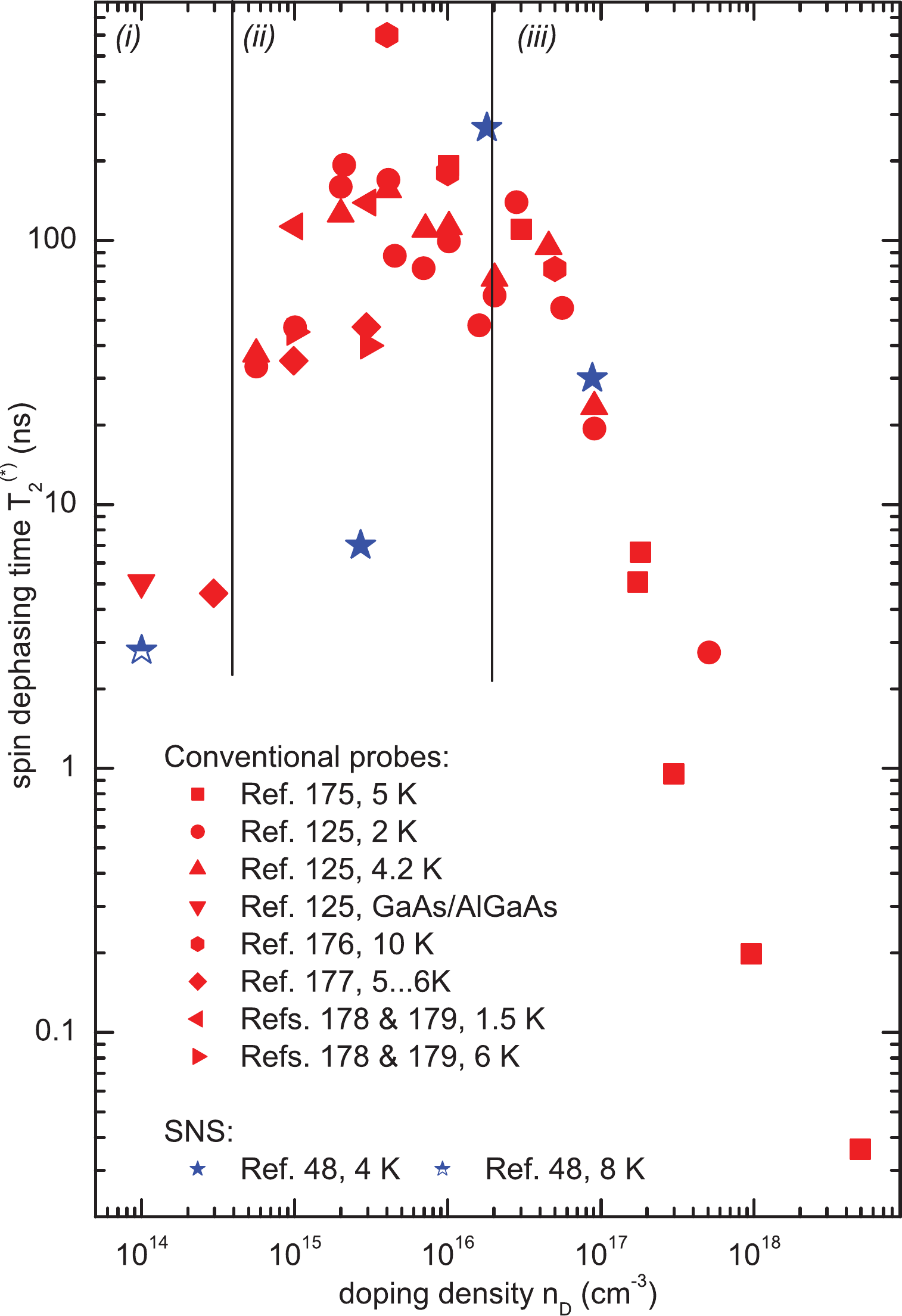}
    \caption{Low temperature spin dephasing times as a function of doping density. Measurements by SNS are indicated by asterisks.  All other data  is acquired by means of resonant spin amplification and time-resolved Faraday rotation \cite{awschalom:physe:10:1:2001},  Hanle measurements \cite{dzhioev:prb:66:245204:2002,furis:njp:9:347:2007}, optically detected electron spin resonance \cite{colton:ssc:132:613:2004}, and time-resolved photoluminescence \cite{colton:prb:69:121307:2004,colton:prb:75:205201:2007}; the data point at $10^{14}\,\mathrm{cm^{-3}}$ of Ref. \cite{dzhioev:prb:66:245204:2002} is measured in a $0.1\,\mathrm{\mu m}$ thick buffer layer of a GaAs/AlGaAs stack.}
    \label{fig:joef}
\end{figure}
SNS was utilized  to study the temperature \cite{romer:rsi:78:103903:2007, crooker:prb:79:035208:2009, romer:prb:81:075216:2010} and transverse magnetic field \cite{mueller:prb:81:121202:2010} dependence of electron spin dephasing in $n$-type bulk GaAs in different doping regimes. Also, spatially-resolved measurements of spin dynamics that are feasible via SNS for all three spatial dimensions (see Sec. \ref{applications:spatial}) delivered a better understanding of inhomogeneous spin dephasing mechanisms for doping regimes close to the metal-to-insulator transition  \cite{mueller:prb:81:121202:2010}.

Figure~\ref{fig:joef} summarizes measured low temperature spin dephasing times as a function of the dopant density, following Fig.~3 of Ref. \cite{dzhioev:prb:66:245204:2002}. The plotted data is acquired via SNS and  various other experimental techniques---like Hanle type measurements, time resolved experiments, resonant spin amplification, and optically detected electron spin resonance (see Sec.~\ref{spindyn:conventional}). The spin dephasing in bulk GaAs can be divided into three regimes:  \textit{(i)}  At very low densities of $n_{\mathrm{D}}\lesssim 10^{14}\,\mathrm{cm^{-3}}$,  all donor electrons can be viewed as non-interacting and the ensemble spin dephasing time is determined by the inhomogeneous distribution of nuclear fields [see Eq.~(\ref{eq:hfnomotional})]. SNS has delivered an inhomogeneous spin lifetime of $T_2^{\ast}=2.8(7)\,\mathrm{ns}$ \cite{romer:prb:81:075216:2010} which is very close to the theoretical value of ${T_2^{\mathrm{HF \ II}}}^{\ast}=3.6\,\mathrm{ns}$ \cite{merkulov:prb:65:205309:2002}. The presence of free carriers in other experimental techniques leads to averaging of the nuclear fields and, hence, conventional measurements can only deliver an upper bound of the spin dephasing time.  \textit{(ii)} With augmenting doping densities, the exchange interaction yields enhanced motional narrowing of the hyperfine interaction induced spin dephasing [see Eq.~(\ref{eq:hfmotional})] and the spin dephasing times increase. Again, the spin dephasing time at $n_{\mathrm{D}}=2.7\times 10^{15}\, \mathrm{cm^{-3}}$ \cite{romer:prb:81:075216:2010} acquired via SNS is significantly smaller than the values that are measured via the various other techniques in this doping regime, ranging from 40 to 600~ns. It is quite probable that this significant deviation does not result from sample specifics like the exact doping regime or the degree of compensation but again from avoiding excitation of free electrons which mitigate spin dephasing by nuclear fields via exchange averaging. The corresponding correlation times that can be acquired in Hanle-like measurements in a longitudinal field \cite{dzhioev:prb:66:245204:2002}   are several orders of magnitude smaller than theoretically expected and rather correspond to values expected in the case of interaction with free electrons \cite{kavokin:sst:23:114009:2008} which explains the more efficient motional narrowing  [see Eq.~(\ref{eq:hfmotional})]. Further, the strong temperature dependence reported in Refs.~\cite{colton:prb:69:121307:2004} and \cite{colton:prb:75:205201:2007} is not found via SNS \cite{romer:prb:81:075216:2010} and may also be an indication of insufficient cooling of the carrier system. With increasing doping density, spin dephasing originating from hyperfine interaction becomes more and more inefficient while the  processes based on anisotropic exchange interaction (see Sec.~\ref{spindyn:deph}) and possibly other processes become more effective. 
At the metal-to-insulator transition ($n_{\mathrm{D}}^{\mathrm{MIT}}=1...2\times 10^{16}\, \mathrm{cm^{-3}}$) low temperature spin dephasing times attain their maximum. The efficiency of the various mechanisms of spin dephasing at the metal-to-insulator transition  has been debated recently \cite{kavokin:sst:23:114009:2008,shklovskii:prb:73:193201:2006,gorkov:prb:67:033203:2003,  tamborenea:prb:76:085209:2007} since the spin dephasing times observed in conventional experiments seem to be too low to be explained by the known mechanisms. Nonetheless, the value acquired by means of SNS of $T_2=267\,\mathrm{ns}$ at $n_{\mathrm{D}}=1.6\times 10^{16}\, \mathrm{cm^{-3}}$ \cite{romer:prb:81:075216:2010} fits well to the theoretical values around $300\,\mathrm{ns}$ that were put forward by Gorkov and Krotkov \cite{gorkov:prb:67:033203:2003}.  
\begin{figure}[tb!]
    \centering
        \includegraphics[width=1.00\columnwidth]{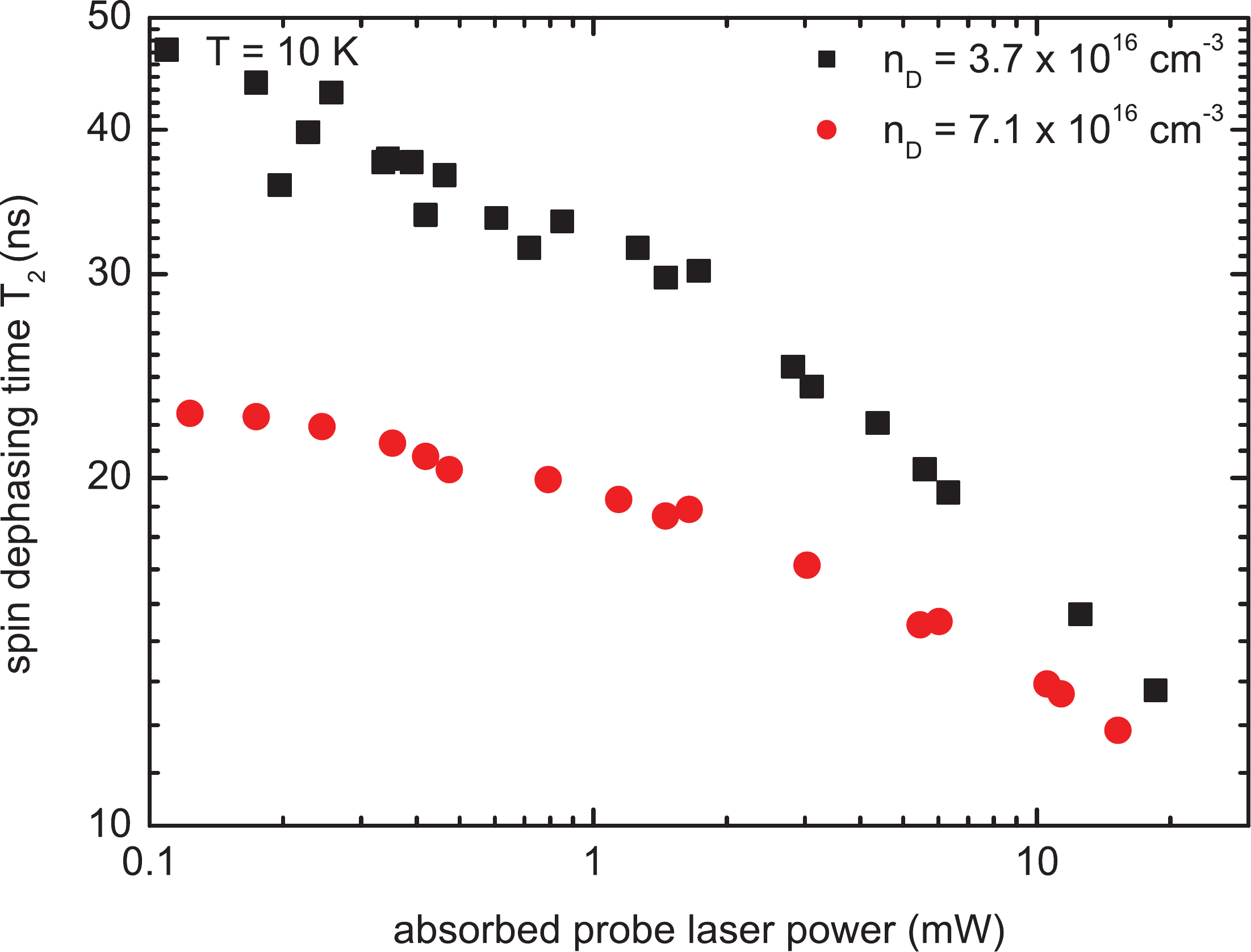}
    \caption{Spin dephasing time measured by spin noise spectroscopy in two $n$-type bulk GaAs samples above the metal-to-insulator transition as a function of the absorbed probe laser power. The amount of absorbed probe laser power is varied either by tuning the laser power or the laser wavelength from 827~nm to 850~nm.  Data is taken from Ref.~\cite{crooker:prb:79:035208:2009}.}
    \label{fig:crooker}
\end{figure}
In this doping regime, the energy deposition in the sample that necessarily accompanies conventional experimental probes obviously reduces the measured spin dephasing times. This assertion is backed up by Fig.~\ref{fig:crooker} which displays the spin dephasing time measured by SNS in two $n$-type bulk GaAs samples above the metal-to-insulator transition as a function of the laser power that is deposited in the sample. The sample with a dopant concentration of $n_{\mathrm{D}}=3.7\times 10^{16}\, \mathrm{cm^{-3}}$, only slightly above the metal-to-insulator transition,  shows a very drastic dependence on the amount of absorbed laser power that does not converge even for the lowest tested values of deposited probe laser power. Accordingly, long spin lifetimes  in bulk GaAs at the metal-to-insulator transition as given in Ref.~\cite{romer:prb:81:075216:2010}  can only be acquired by means of SNS with strong detuning from the resonance as well as increased probe laser spot size.
 \textit{(iii)} In the metallic regime, at doping densities above the metal-to-insulator transition, spin dephasing is predominantly determined by the DP process.  The electronic energy and, hence, the spin splitting relevant for the efficiency of the DP process increases with higher doping densities [see Eq.~(\ref{eq:dpgaas})].   Sample excitations due to optical orientation play a less important role in this degenerate doping regime (see Sec. \ref{spindyn:conventional}). Therefore all experimental techniques deliver similar results in this regime. The data for the sample with $n_{\mathrm{D}}=7.1\times 10^{16}\, \mathrm{cm^{-3}}$ in Fig.~\ref{fig:crooker} proves this reasoning since the spin dephasing time is signifcantly less dependent on the amount of absorbed laser power than at lower doping intensities.

\begin{figure}[tb!]
    \centering
        \includegraphics[width=1.00\columnwidth]{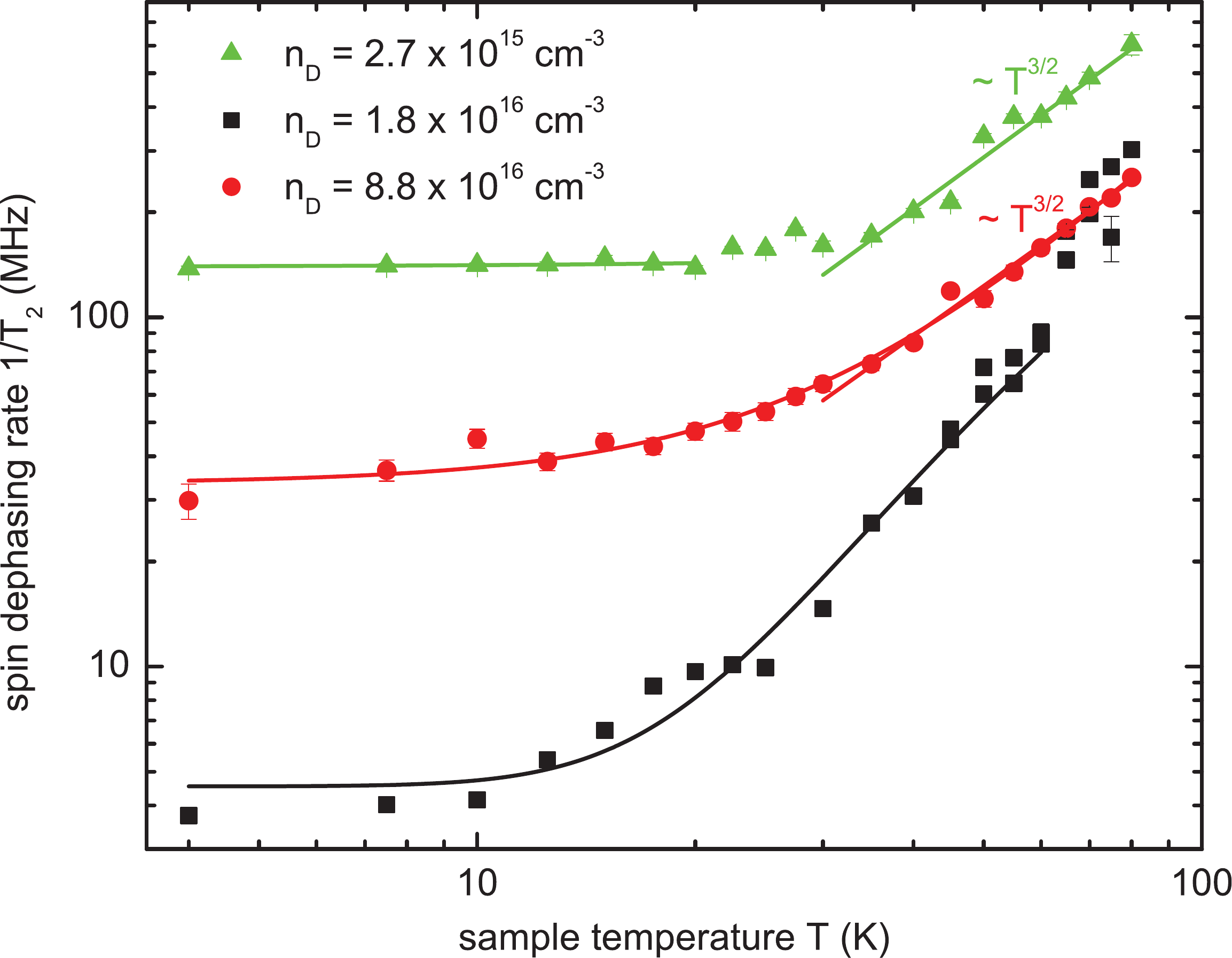}
    \caption{Temperature dependence of the spin dephasing rates in $n$-type bulk GaAs for different doping densities. All data is acquired via SNS and taken from Ref.~\cite{romer:prb:81:075216:2010}.}
    \label{fig:temp}
\end{figure}
The equilibrium sample temperature is a well defined experimental parameter in SNS since carrier heating is avoided. Thus, SNS is an ideal tool to study the temperature dependence of the spin dephasing rates in the different doping regimes. Spin dephasing times in $n$-type GaAs ranging from liquid helium temperatures up to 80~K are given in Ref.~\cite{romer:prb:81:075216:2010} and are reproduced in Fig.~\ref{fig:temp}. The spin dephasing  at low temperatures is found to depend only weakly on temperature or---in the case of the investigated sample with the lowest doping concentration---is even independent of temperature ($n_{\mathrm{D}}=2.7\times 10^{15}\, \mathrm{cm^{-3}}$). In the sample at the metal-to-insulator transition ($n_{\mathrm{D}}=1.8\times 10^{16}\, \mathrm{cm^{-3}}$), the electrical conductivity 
shows a very similar temperature behavior as the  spin dephasing rate, indicating a quite direct relation between the spatial electron dynamics and spin dephasing \cite{romer:prb:81:075216:2010}.  A substantial amount of the donor atoms in low doped samples is ionized  at elevated temperatures  and spin dephasing is governed by the DP process for all doping concentrations. Here, conventional experimental methods are expected to be equally sensitive as SNS. In any case, the data in Fig.~\ref{fig:temp} reveals that at elevated temperatures the spin lifetimes are no more the longest at the metal-to-insulator transition, but at higher doping concentrations where motional narrowing via momentum scattering at impurity atoms is more efficient [see Eq. (\ref{eq:dpgaas})]. Scattering at ionized impurities further becomes manifest in the $T^{3/2}$ temperature dependence of the dephasing rates \cite{zutic:rmp:76:32:2004}  which is observed in Fig.~\ref{fig:temp}. The doping and temperature dependence of $n$-type bulk GaAs is comprehensively studied in Ref.~\cite{jiang:prb:79:125206:2010} by means of  fully microscopic calculations which further include electron-electron and electron-phonon scattering. These theoretical studies give a more detailed temperature  dependence as indicated by the fits in Fig.~\ref{fig:temp}. 

\begin{figure}[tb!]
    \centering
        \includegraphics[width=1.00\columnwidth]{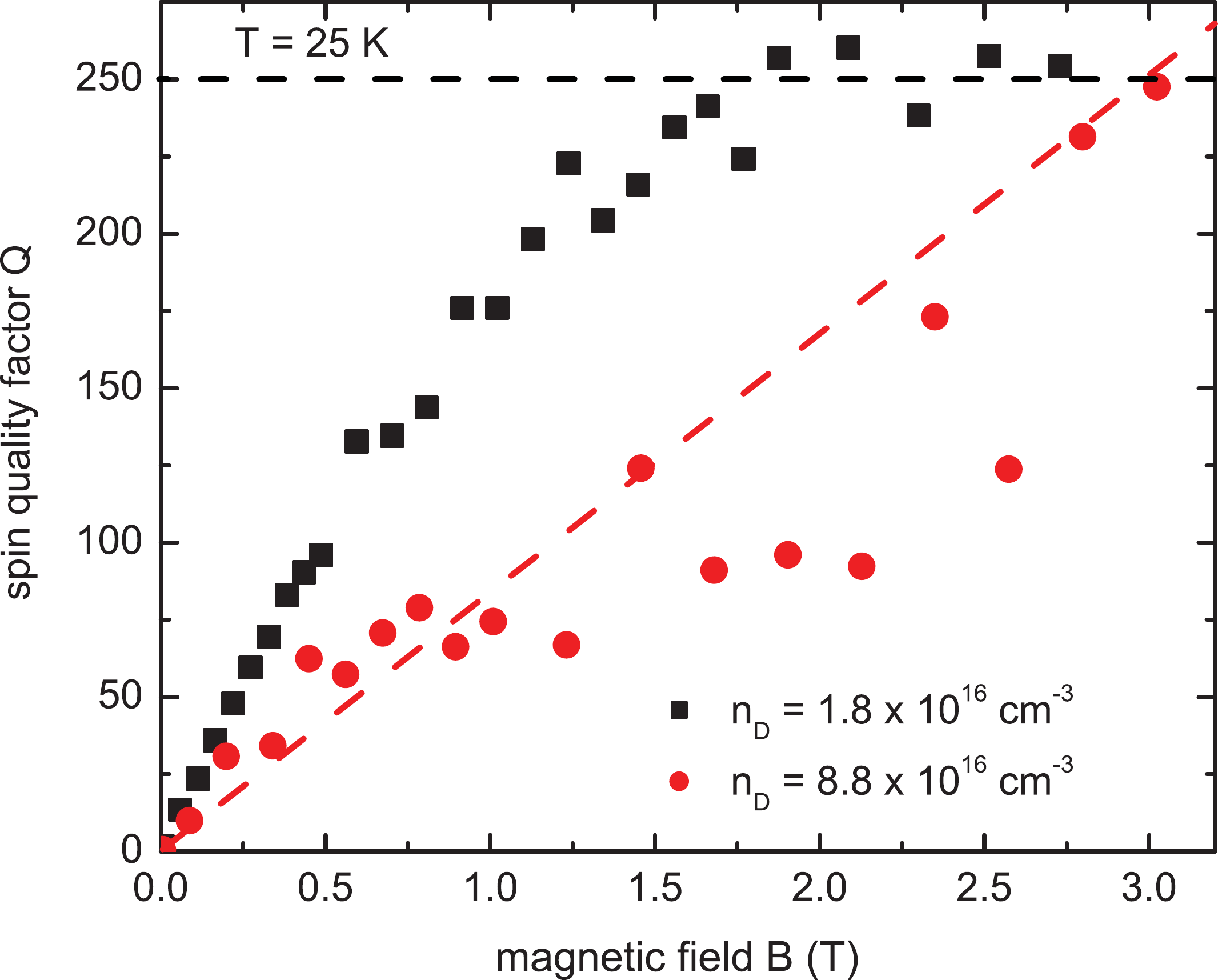}
    \caption{Spin quality factor $Q=g^\ast\mu_{\mathrm{B}}BT^\ast_2/h$ as a function of the transverse magnetic field in $n$-type bulk GaAs for different doping densities measured via SNS. The lines are guides to the eye. The data is reproduced from Ref.~\cite{mueller:prb:81:121202:2010}.}
    \label{fig:quality}
\end{figure}

A spread of the effective $g$-factor in the sample may  yield an inhomogeneous broadening of the spin dephasing  in high transverse magnetic fields \cite{kikkawa:prL:80:43131998,wu:epjb:18:373:2000,putikka:prb:70:113201:2004,margulis:spss:25:928:1983,bronold:prb:66:2332062002}.   Such effects can be investigated via SNS because of the  recently achieved  advancement of SNS to GHz frequencies (see. Sec.~\ref{experiment:ultrafast}). M\"uller \textit{et al.} examined the spin dynamics in high transverse magnetic fields in two $n$-type bulk GaAs samples: one very close at the metal-to-insulator transition ($n_{\mathrm{D}}=1.8\times 10^{16}\, \mathrm{cm^{-3}}$), the other well above  ($n_{\mathrm{D}}=8.8\times 10^{16}\, \mathrm{cm^{-3}}$) \cite{mueller:prb:81:121202:2010}. Spin dephasing in high magnetic fields is quantitatively well characterized by means of the spin quality factor $Q=g^\ast\mu_{\mathrm{B}}BT^\ast_2/h$  \cite{kikkawa:prL:80:43131998} which is plotted as a function of the applied magnetic field in Fig.~\ref{fig:quality}: In the metallic doping regime, the $Q$-factor increases with the applied field and, hence, no inhomogeneous broadening is observed in accord with existing studies \cite{kikkawa:prL:80:43131998} and the theoretical expectation that delocalized electronic states average out all inhomogeneities \cite{bronold:prb:66:2332062002}. The inhomogeneous broadening close to the metal-to-insulator transition, which becomes manifest  in a transition from a Lorentzian to a Gaussian line shape in the spin noise spectra (see Fig.~\ref{fig:ghz}) as well as in the formation of a $Q$-factor plateau, is  around a factor of three less pronounced than in a similar sample examined by resonant spin amplification \cite{kikkawa:prL:80:43131998}. The higher temperature used in the SNS experiments of 25~K compared to the resonant spin amplification experiment at liquid helium temperatures directly disproves the assertion that the inhomogeneous broadening in these cases results from a  spread of the effective electronic $g$-factor due to a thermal spread of the electronic energy. Instead, a spatial $g$-factor variation is found in the investigated sample \cite{mueller:prb:81:121202:2010} measured by SNS with spatial depth resolution (see Sec.~\ref{applications:spatial}): The absolute value of the $g$-factor is increased at the sample surfaces which may result  from surface depletion. Delocalized electrons would average over such spatial inhomogeneities and no increase of the spin dephasing rate would be observed, but SNS directly reveals that the electrons in the investigated sample ($n_{\mathrm{D}}=1.6\times 10^{16}\, \mathrm{cm^{-3}}$) are to some extent localized \cite{mueller:prb:81:121202:2010}. This can be deduced from the temperature dependence of the observed spin noise power as already mentioned in Sec.~\ref{sns:deloc}. 

In general, the temperature dependence of the spin noise power can be divided in three distinct regimes, of which all are found in the experiment \cite{romer:prb:81:075216:2010}.  An extrapolation to zero temperature should deliver vanishing spin noise power in the case of a degenerate electron gas in which spin flips are suppressed at 0~K due to the Pauli principle. For localized electrons, the spin noise power is independent of the temperature and in the intermediate doping regime close to the metal-to-insulator transition, where electron transport proceeds via hopping,   a mixed behavior with residual spin noise power at zero temperature is found,  proving partial localization of electrons.

\paragraph{GaAs/AlGaAs quantum wells}
\begin{figure}[tb!]
    \centering
        \includegraphics[width=1.00\columnwidth]{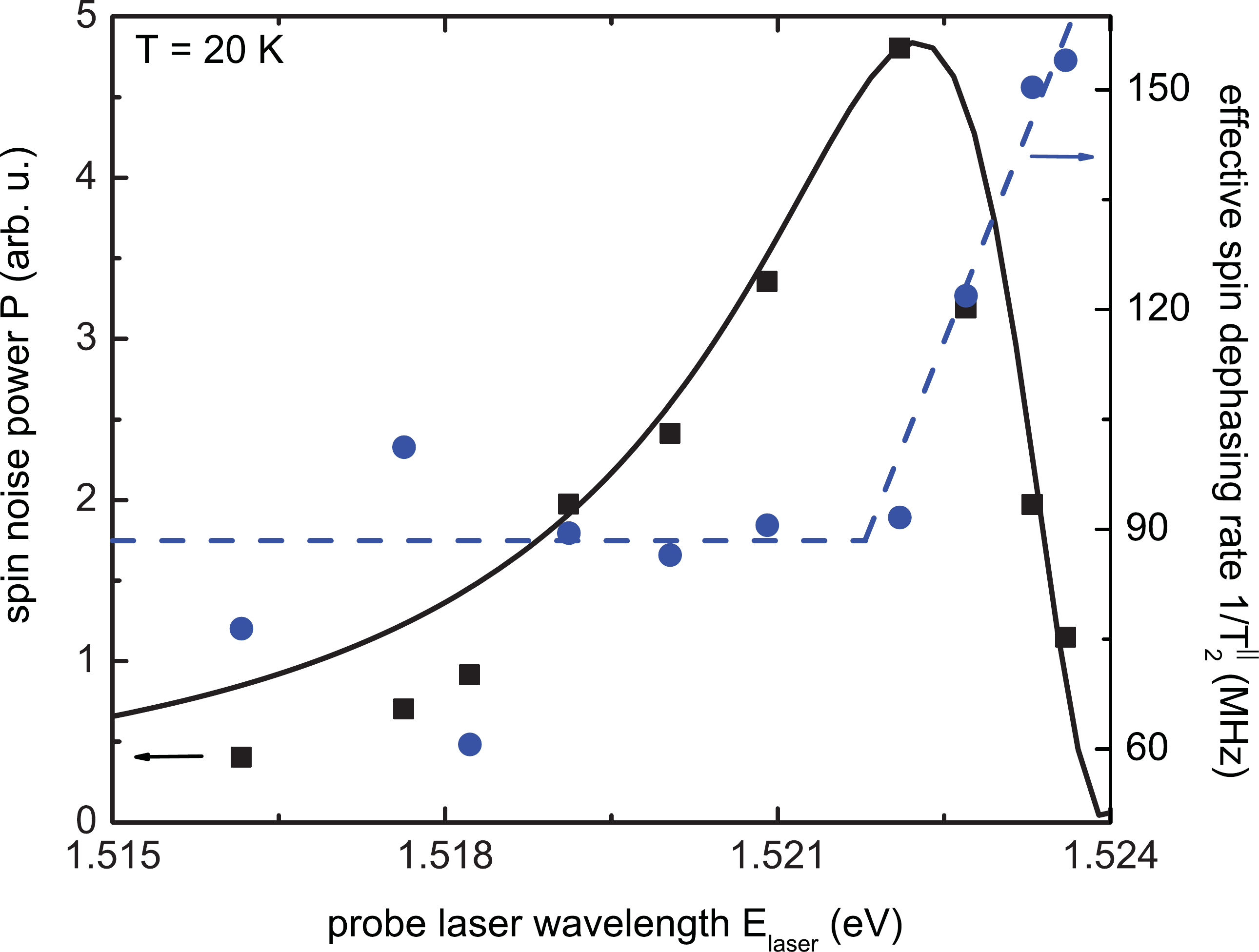}
    \caption{Spin noise power $P$ and effective spin dephasing rate $1/T_2^{\|}$ measured in an (110) grown GaAs/AlGaAs multiple quantum well structure (each quantum well: $n_{\mathrm{D}}=1.8\times 10 ^{11}\,\mathrm{cm^{-2}}$, thickness 16.8~nm) as a function of the probe laser energy at $T=20\,\mathrm{K}$. The solid line is a calculation of the spin noise power  according to the modeling described in Sec.~\ref{sns:deloc}.  As expected from Eq.~(\ref{eq:noisepower}), the data resembles nicely the square of the real part of a refractive index  within the Lorentz oscillator model (see Fig.~\ref{fig:fig2-absorption-refraction}). The spin dephasing rate increases significantly by tuning to the resonance due to optical creation of holes. Time of flight effects also contribute to the measured spin dephasing rate. The dashed curve is a guide to the eye. All data is taken from Ref.~\cite{muller:prl:101:206601:2008}.}
    \label{fig:qw}
\end{figure}
Semiconductor quantum wells attract a lot of attention  in the context of spintronics since they  allow to tailor spin orbit fields (see Sec.~\ref{spindyn:deph}).   GaAs based quantum wells with an $(110)$ growth axis are especially interesting since the Dresselhaus field points along the growth axis for all $\mathbf{k}$-states such that electronic spins aligned with the growth axis do not  dephase according to the DP mechanism \cite{dyakonov:ssps:20:110:1986}. The longer spin dephasing times in (110) grown quantum wells compared to equivalent (001) structures were experimentally shown by Ohno \textit{et al.} in 1999 \cite{ohno:prl:83:4196:1999}. Later, D\"ohrmann and co-workers demonstrated that spins in the quantum well plane still undergo spin dephasing via the DP process and that, subsequently, spin dephasing is anisotropic in (110) grown structures \cite{dohrmann:prl:93:147405:2004}. However, in this investigation via time and polarization resolved photoluminescence, the anisotropy of the spin dephasing is diminished at low temperatures due to the BAP mechanism. Like in all experimental probes  that rely on optical spin orientation (see Sec. \ref{spindyn:conventional}), the presence of optically created holes yields additional spin dephasing  which becomes dominant because of the enhanced exchange interaction at low temperatures and the absence of other efficient mechanisms of spin dephasing. In 2007, Couto \textit{et al.} spatially separated \cite{couto:prl:98:036603:2007} optically created holes from electrons by means of surface acoustic waves. Nevertheless, the influence of these acoustic waves to the spin dephasing had not been established yet  and, hence,  the dominant process of spin dephasing  and the corresponding spin lifetimes  in (110) GaAs quantum wells at  low temperatures remained unknown.

\begin{figure}[tb!]
    \centering
        \includegraphics[width=1.00\columnwidth]{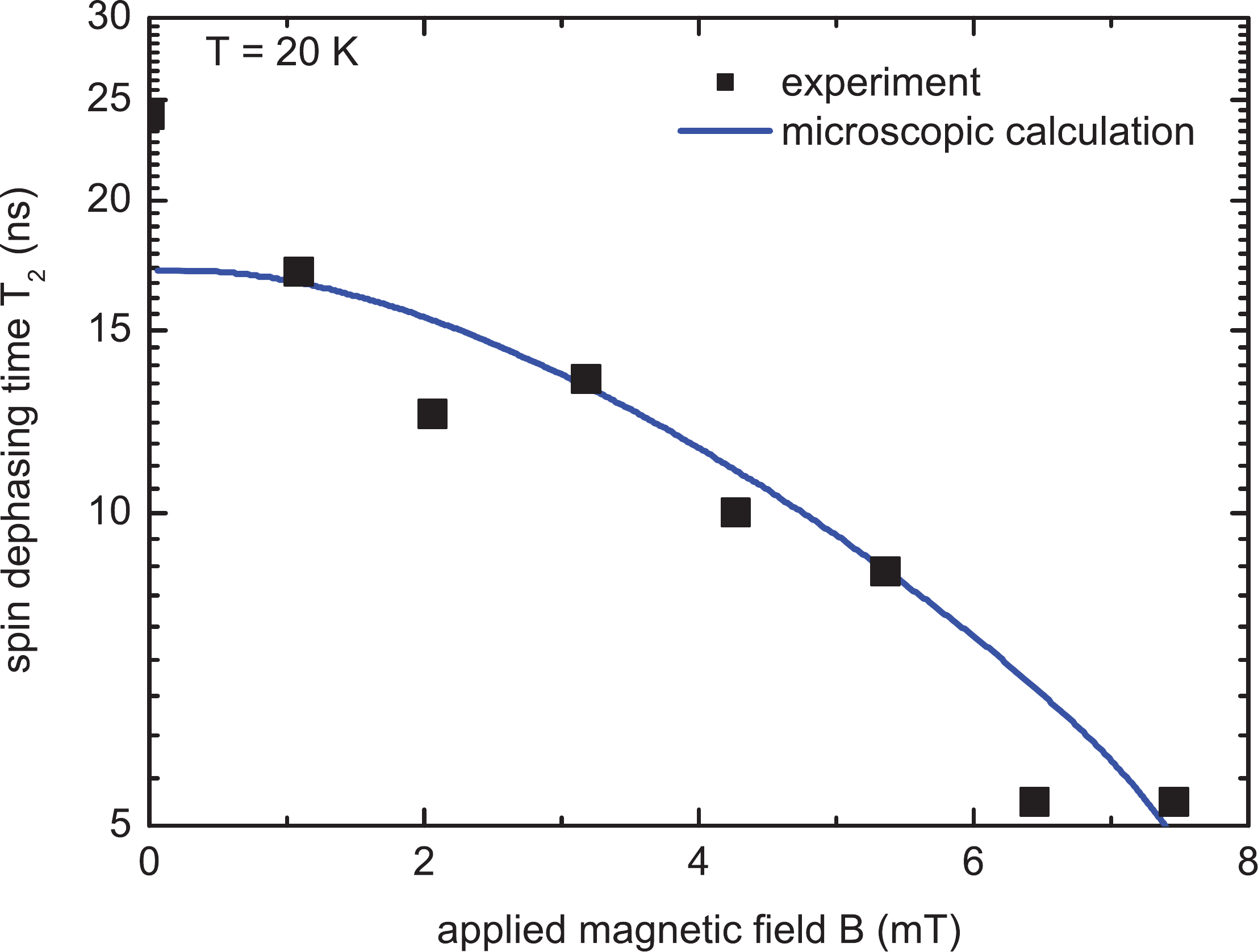}
    \caption{Spin dephasing time $T_2$ measured in an (110) grown GaAs/AlGaAs multiple quantum well structure (each quantum well: $n_{\mathrm{D}}=1.8\times 10 ^{11}\,\mathrm{cm^{-2}}$, thickness 16.8~nm) as a function of an applied in-plane magnetic field at $T=20\,\mathrm{K}$. The magnetic field rotates the spins aligned along the growth axis into the quantum well plane and, hence, they are subject of spin dephasing according to the DP mechanism. The line represents the spin dephasing time calculated by means of kinetic spin Bloch equations where random spin orbit fields are additionally taken into account. The experimental data is taken  from Ref.~\cite{muller:prl:101:206601:2008} and the theory curve  from Ref.~\cite{zhou:epl:89:57001:2010}.}
    \label{fig:qw2}
\end{figure}
In 2009,  application of SNS to a modulation doped (110) grown GaAs/AlGaAs multiple quantum well structure (each quantum well: $n_{\mathrm{D}}=1.8\times 10 ^{11}\,\mathrm{cm^{-2}}$, thickness 16.8~nm) \cite{muller:prl:101:206601:2008} enabled the investigation of electron spin dephasing at low temperatures in the absence of optically generated electron-hole pairs. The potentially strong influence of the BAP process becomes manifest in a drastic increase of the spin dephasing rates by tuning the probe laser close to the optical transition (see Fig.~\ref{fig:qw}) while the measured spin noise power resembles  the square of the real part of a refractive index  within the Lorentz oscillator model (see Fig.~\ref{fig:fig2-absorption-refraction})  as described in Sec.~\ref{sns:deloc}. In this SNS experiment, also the finite transit times of the probed electrons through the laser spot play an important role. However, with an enlarged laser spot, to avoid these time of flight effects, and with the laser detuned from the resonance, to avoid excitation of holes, SNS delivers the intrinsic spin dephasing times of the investigated sample structure. A spin dephasing time for spins aligned along growth axis of $T_2^{\|}=24(2)\,\mathrm{ns}$ is measured at 20~K. The anisotropic spin dephasing that is  concealed in time and polarization-resolved photoluminescence experiments  at these temperatures \cite{dohrmann:prl:93:147405:2004} is also recovered via SNS, where a ratio of $T_2^{\|}/T_2^{\perp}=7.4(1.0)$ between the dephasing times of spins aligned along and perpendicular to the growth direction is measured by application of an in-plane magnetic field (see Fig.~\ref{fig:qw2}). The observed  lifetimes $T_2^{\|}$ cannot be limited by one of the well studied spin dephasing mechanisms   (Sec.~\ref{spindyn:deph}). Also, the recently discovered intersubband spin relaxation \cite{dohrmann:prl:93:147405:2004,hagele:assp:45:253:2006} cannot completely account for the experimental findings  according to the microscopic calculations by Zhou and Wu \cite{zhou:ssc:149:2078:2009}.   M\"uller \textit{et al.} \cite{muller:prl:101:206601:2008} suggested that the observed  lifetimes $T_2^{\|}$ are limited  by a mechanism that was initially put forward by Sherman in 2003 \cite{sherman:apl:82:209:2003} which results from random spin-orbit fields arising from electrical fields due to inevitable spatial fluctuations of the impurity atoms in the $\delta$-doping sheets (see also Ref.~\cite{GlazovPE2010}). Recently, several theoretical investigations on spin dephasing due to random spin-orbit fields as well as on spin dynamics in (110) grown GaAs quantum wells were published \cite{zhou:ssc:149:2078:2009,zhou:epl:89:57001:2010,dugaev:prb:80:081301:2009, tarasenko:prb:80:165317:2009, glazov:prb:81:115332:2010, tokatly:aop:325:1104:2010}. While the microscopic calculation from Ref.~\cite{zhou:epl:89:57001:2010} agrees well with the experimental findings, especially with the measured magnetic field dependence (see Fig.~\ref{fig:qw2}), the work by Glazov \textit{et al.}, where also spin-flip collisions of electrons from different quantum wells of the multiple quantum well structure are explicitly considered \cite{glazov:prb:81:115332:2010}, implies that still additional,  even unknown processes may contribute to the observed spin dephasing. The aforementioned transit time effects obviously pose a challenge for acquiring the intrinsic spin lifetimes. However, this time of flight broadening also implicates a great potential since it  uniquely allows to study spatial electron dynamics at thermal equilibrium \cite{muller:prl:101:206601:2008}.

\paragraph{(In,Ga)As/GaAs quantum dots}
\begin{figure}[tb!]
    \centering
        \includegraphics[width=1.00\columnwidth]{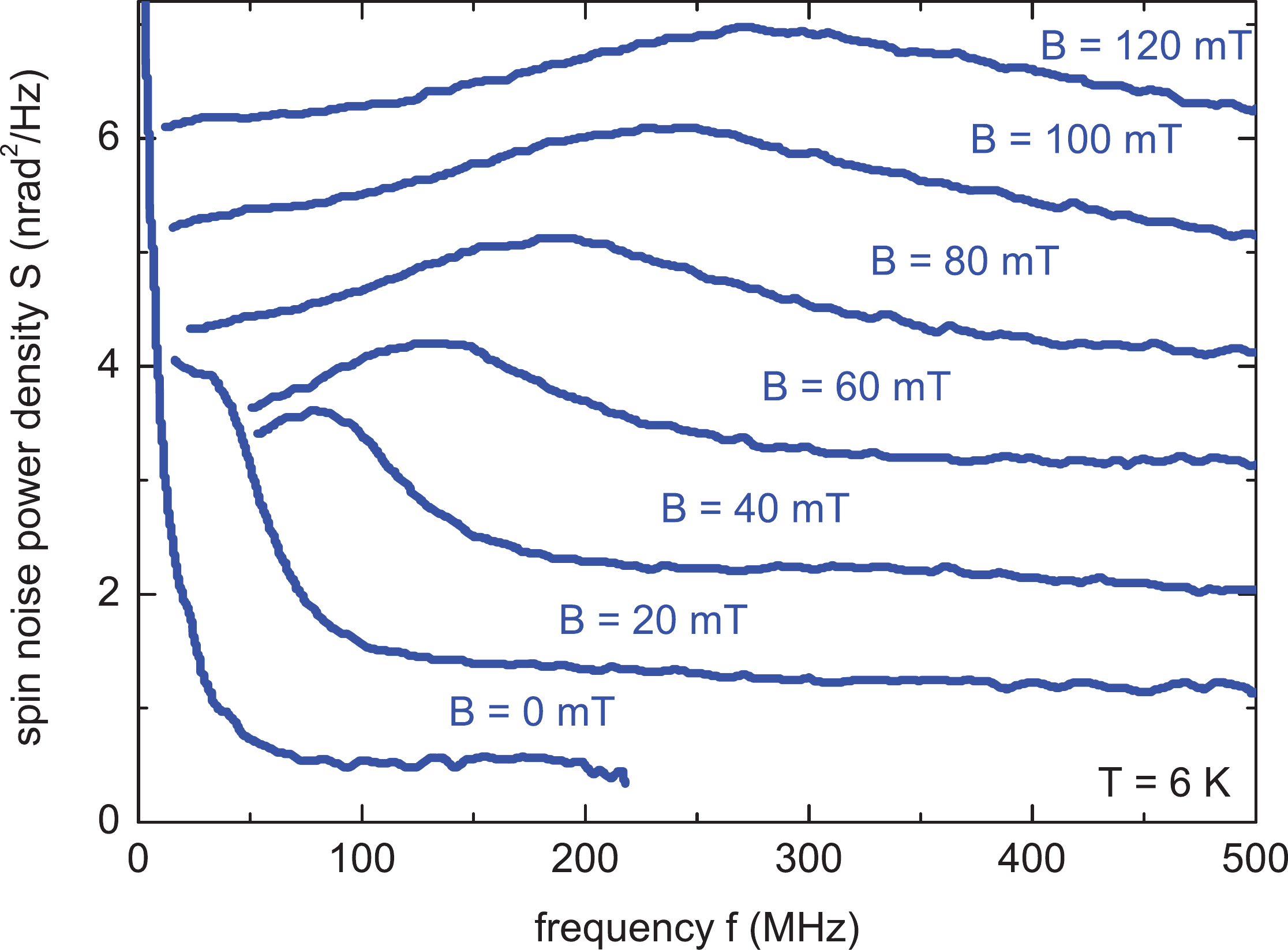}
    \caption{Spin noise spectra of holes confined in self-assembled (In,Ga)As/GaAs quantum dots. A strong inhomogeneous broadening in a transverse magnetic field is observed. The spectra are shifted for clarity.  Data is taken from Ref.~\cite{crooker:prl:104:036601:2010}.}
    \label{fig:qd}
\end{figure}
The recent work by Crooker \textit{et al.} \cite{crooker:prl:104:036601:2010} represents a compelling proof-of-principle experiment revealing that SNS is by far sensitive enough to detect spin dynamics of electrons and holes effectively confined to zero dimensions. Besides the first SNS experiment on self-assembled quantum dots, Ref.~\cite{crooker:prl:104:036601:2010} also contains the first spin noise measurements of hole spins and by above bandgap illumination intentionally optically created electrons, of which both are neatly identified by the specifics of the effective $g$-factor. The investigated sample structure consists of 20 layers of (In,Ga)As/GaAs grown by molecular beam epitaxy on a (100) GaAs substrate where each layer has a quantum dot density of around $10^{10}\,\mathrm{cm^{-2}}$. The relatively large number of probed quantum dots  and the large inhomogeneous spread of the confinement energy around $0.2\,\mathrm{eV}$, which also results in the strong broadening of the spin noise curves in high transverse magnetic fields (see Fig.~\ref{fig:qd}),   preclude  demolition free application of SNS by detuning from the probed resonance. Still, this experiment may pave the way for application of SNS on single quantum dots.  The detection of a single electron spin by below band gap Faraday \cite{atature:natphys:3:101:2007} and Kerr \cite{berezovsky:science:324:1916:2006, mikkelsen:natphys:3:770:2007} rotation has already been established which shows that SNS of a single electron spin should be feasible.

%% file: Review_Paper_Section456_002.tex
\section{Experimental Aspects of SNS}\label{experiment}
Over the last years, semiconductor SNS has developed into a very sensitive tool to study spin dynamics in semiconductors. The sensitivity has significantly increased and reached a level that allows to apply SNS to quantum wells \cite{muller:prl:101:206601:2008}, quantum dot arrays \cite{crooker:prl:104:036601:2010}, and even only a few microns thick epilayers of bulk semiconductor material \cite{romer:prb:81:075216:2010}. Recently, the technical limitation of SNS to frequencies within the bandwidth of the balanced photoreceiver has been overcome and SNS was demonstrated at frequencies of several GHz \cite{mueller:prb:81:121202:2010}. This section is devoted to rather technical aspects that are only parenthetically mentioned in the corresponding research papers, but are crucial for the achieved advancements. In Sec.~\ref{experiment:substract}, several possibilities to separate the actual spin noise from other noise contributions are discussed. Efficient data averaging, which is of great importance to semiconductor SNS, is discussed in Sec.~\ref{experiment:averaging}. In Sec.~\ref{experiment:ultrafast}, the rather new advancement of SNS to GHz frequencies is presented.

\subsection{Shot Noise Subtraction}\label{experiment:substract}
Spin noise is not the only noise contribution that is detected in SNS. While classical noise is eliminated by stable laser sources and balanced detection, optical shot noise is always present and exceeds the amount of spin noise by several orders of magnitude  as discussed in Sec.~\ref{sns:spurious}. Additionally, commercial detectors with the necessary bandwidth for semiconductor SNS of 100~MHz to 1~GHz exhibit electrical noise that is not negligible at low probe powers. Laser shot noise is white noise and usually adds as a constant noise floor to the spin noise. However, the frequency response of the detector and an optional pre-amplifier can generally not be viewed as constant within the frequency intervals given by the spin noise width.  Subtraction of the background noise floor is therefore necessary to avoid distortion of the spin noise spectra. To this end, a reference noise curve that does not contain spin noise has to be acquired. This can be achieved by shifting the spin noise peak in frequency  by  variation of the applied magnetic field as demonstrated in Fig.~\ref{fig:fig5-lorentz} and in Refs.~\cite{oestreich:prl:95:216603:2005,romer:rsi:78:103903:2007,crooker:prb:79:035208:2009, crooker:prl:104:036601:2010,mueller:prb:81:121202:2010}.

Alternatively, a reference noise spectrum can be acquired by switching the optical bridge setup from detection of circular birefringence, i.e., Faraday rotation, to linear birefringence  and thereby suppressing the spin noise signal contained in the probe light, while  keeping the photon shot noise background. 
The suppression of spin noise can be achieved by two distinct schemata: \textit{(i)} The polarization state of the probe laser light can be changed from linearly polarized light to circularly polarized light before it is transmitted through the sample \cite{muller:prl:101:206601:2008}. Here, the circularly polarized light does not acquire a Faraday rotation and  is split in equal parts into the two  orthogonal linear polarization states  via the polarizing beam splitter cube in front of the detector  (see Fig.~\ref{fig:fig4-setup}). This scheme, however, has some disadvantages, e.g., if the sample exhibits linear dichroism due to strain or  magnetic effects. \textit{(ii)} The second scheme eliminates the acquired Faraday rotation  behind the sample  \cite{roemer:apl:94:112105:2009, romer:prb:81:075216:2010}. To this end, the fast axis of a variable retarder behind the sample is aligned along the linear light polarization and the retardation is switched from $\lambda/2$ (no change) to $\lambda/4$ (suppression of spin noise). The variable retardation can be either implemented by  a motorized Soleil-Babinet compensator \cite{muller:prl:101:206601:2008} or a liquid crystal retarder \cite{roemer:apl:94:112105:2009, romer:prb:81:075216:2010}. The usage of the latter is convenient because of the higher switching speed between the two polarization states. Switching the liquid crystal retarder, however, also introduces a slight change of the light transmission due to a change in the absolute refractive index of the waveplate that has to be accounted for in the experiment.


In some cases, best results are achieved if a double difference scheme is utilized by  changing both the light polarization  and the magnetic field \cite{muller:prl:101:206601:2008}. Additionally, the frequency response of the detection has  to be taken into account for reliable measurements of the correct value for the spin noise power.

\subsection{Data Acquisition and Spectrum Analysis}\label{experiment:averaging}
\begin{figure}[tb!]
    \centering
        \includegraphics[width=1.00\columnwidth]{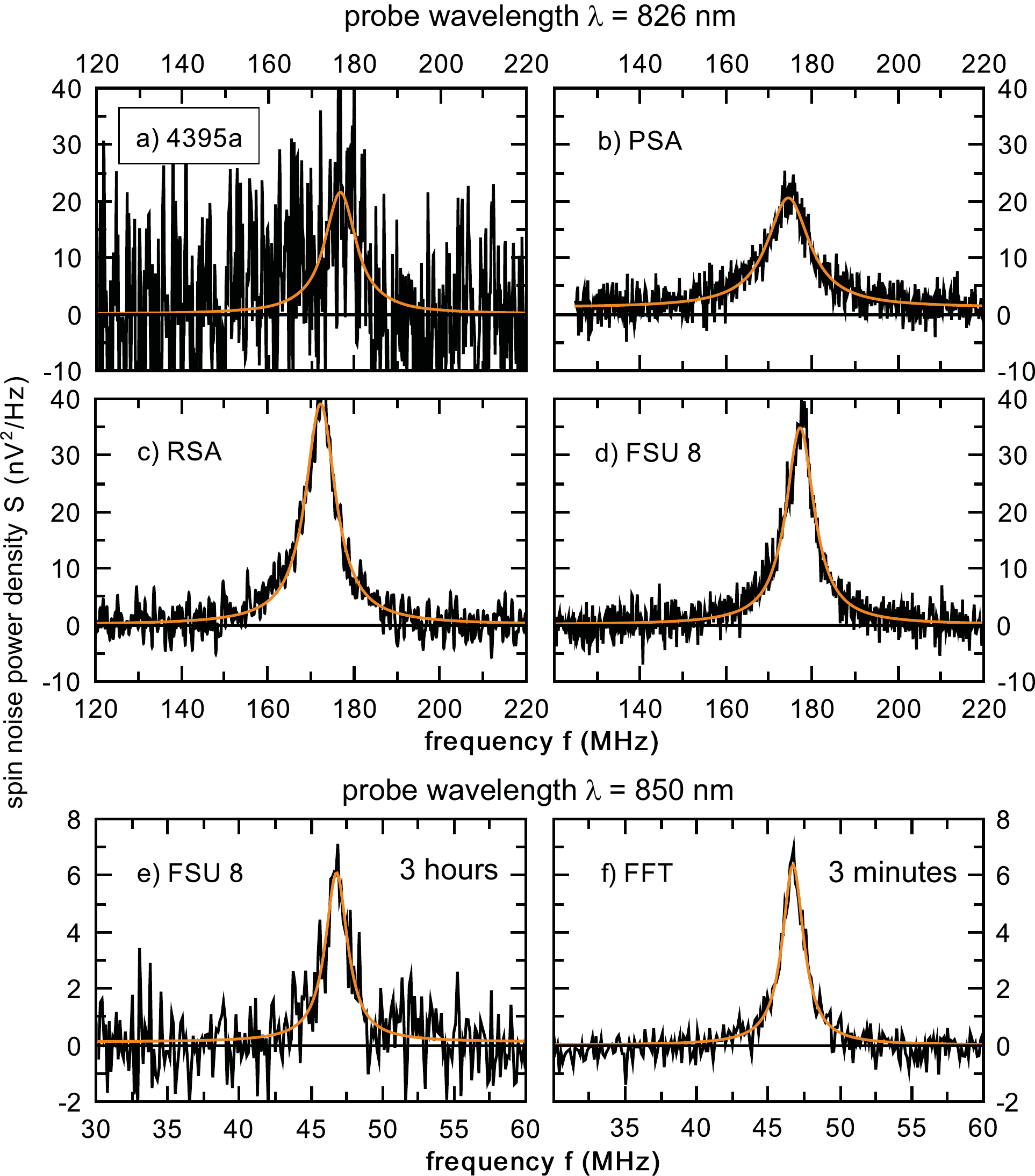}
    \caption{Spin noise spectra ($n$-type GaAs, 10 K) acquired by different spectrum analyzers. (a)-(d) Comparison of commercial analyzers employing a reference oscillator (probe wavelength $\lambda=826\,\mathrm{nm}$, averaging time  4 minutes): (a)  Hewlett \& Packard 4395a, (b) Hewlett \& Packard PSA, (c) Tektronix RSA 3408a, (d) Rhode \& Schwarz FSU 8. (e),(f) Comparison of a sweeping spectrum analyzer (e, Rhode \& Schwarz FSU 8, averaging time  3 hours) with a real-time FFT analyzer (averaging time  3 minutes) at a probe wavelength of $\lambda=850\,\mathrm{nm}$.}
    \label{fig:diffspec}
\end{figure}
In the first paper on semiconductor SNS \cite{oestreich:prl:95:216603:2005}, a sweeping spectrum analyzer was utilized for transforming the acquired time signal into the frequency domain.  Efficient data averaging is of great importance for flattening the shot noise background due to the low ratio of peak spin noise power to background noise density $\eta$ (see Tab.~\ref{tab:SNSdata}). However, a spectrum analyzer with a sweeping local oscillator measures the noise only at the reference frequency  and thereby disregards the majority of the available data stream. Sweeping over 1~GHz bandwidth with a resolution of 1~MHz simply means that around $99.9\%$ of the acquired signal remain unused at a time. Still, different commercial spectrum analyzers show a significant difference in sensitivity as depicted in Figs.~\ref{fig:diffspec}~(a)-(d). Fundamentally, this problem is circumvented by digitizing the data stream and subsequent realtime spectrum analysis via fast Fourier transformation (FFT). The FFT algorithm allows simultaneous detection of spin noise at all frequencies within the detection bandwidth and, hence, with no dead time as long as all digitized data can be further processed, i.e., $100\%$ of the signal acquired in the time domain enter into the data processing and averaging. In order to comprehend the compelling increase of detection sensitivity, Figs.~\ref{fig:diffspec}~(e) and (f) show two SNS spectra acquired by means of a commercial sweeping spectrum analyzer as well as  a FFT spectrum analyzer.   This advance of the SNS setup was first realized by R\"omer \textit{et al.} \cite{romer:rsi:78:103903:2007} and employed in all subsequent publications on semiconductor SNS \cite{muller:prl:101:206601:2008, crooker:prb:79:035208:2009, roemer:apl:94:112105:2009, romer:prb:81:075216:2010, crooker:prl:104:036601:2010, mueller:prb:81:121202:2010}. The actual realtime FFT analysis  is perfectly suitable for parallel computing and, therefore, scalable to high throughput. As the computer's PCI Express bus allows data transmission with rates of up to $16\,\mathrm{GByte/s}$  and multicore CPUs become more and more efficient, software based realtime FFT on the CPU yields an extremely high  data transmission; currently, our group routinely processes the noise signal with a sampling rate of  $f_{\mathrm{S}}= 1$~GSamples/s.  In a similar approach, Crooker \textit{et al.} implemented the   FFT routine  by means of a digitizer incorporating field programmable gate array processors ($f_{\mathrm{S}}= 2$~GSamples/s) \cite{crooker:prl:104:036601:2010}.   According to the Nyquist-Shannon theorem \cite{nyquist:transaiee:47:617:1928, shannon:procire:37:10:1949},  the  SNS setup in  Fig.~\ref{fig:fig4-setup} can only detect spin noise  at frequencies smaller than the detection bandwidth which is given by half of the sampling rate: $B=f_{\mathrm{S}}/2$. It is important to cut off all shot noise at frequencies larger than $B$ by means of low pass frequency filters. Otherwise, undersampling of these frequency components would result in an increased background noise level within the detection bandwidth. 


The bit depth $R$ is another figure of merit for an analog-to-digital converter and specifies together with the sampling rate the data transmission rate of the digitizer $I=f_{\mathrm{S}}\times R$ (see, e.g., Ref.~\cite{JayantNoll1984}). The bit depth determines the quantization error of a digitized signal, i.e., the difference between analog input and digital output. In the case of uniform quantization and avoidance of overload of the digitizer,  the variance of the quantization error reads  $\Delta^{-2}/12$ according to Bennett's famous approximation \cite{bennett:bellstj:27:446:1948}.\footnote{For a discussion of the validity of this approximation, see, e.g., Ref.~\cite{gray:ieeetransit:36:1220:1990}.}  Here, $\Delta\propto 2^{-R}$ gives the size of the least significant bit.  Thus, the variance of the quantization error scales exponentially with the utilized number of bits per sample. Interestingly,  the  signal-to-noise ratio in SNS is not  limited by this quantity: The ever present shot noise floor (see Sec.~\ref{experiment:substract}) represents an additive dither (see, e.g., Refs.~\cite{schuchmann:ieeetranscom:12:162:1964, gray:ieeetransit:39:805:1993, carbone:ieeetransim:43:389:1994, wannamaker:ieeetranssp:48:499:2000, skartlien:ieeetransim:54:103:2005}) to the spin noise signal which facilitates quite efficient averaging of the quantization error. A detailed understanding of the interplay of averaging and quantization errors is necessary to achieve the maximal sensitivity for SNS. An in-depth investigation on the sensitivity of SNS due to quantization errors will be published elsewhere
\cite{mueller:inprep}.

\subsection{GHz Spin Noise Spectroscopy}\label{experiment:ultrafast}
\begin{figure}[tb!]
   \centering
       \includegraphics[width=1.00\columnwidth]{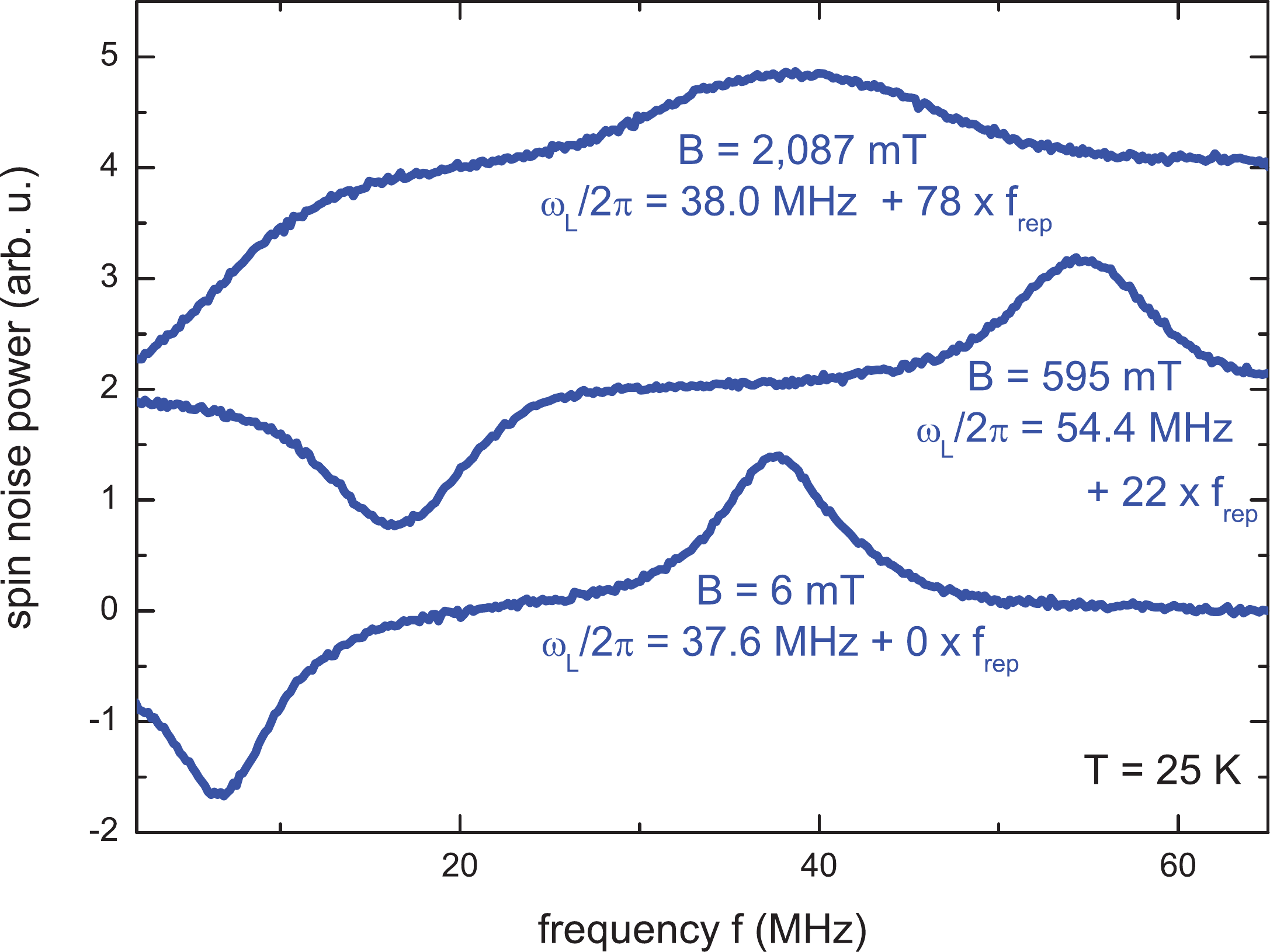}
   \caption{Spin noise spectra acquired by GHz SNS \cite{mueller:prb:81:121202:2010}. The repetition rate of the probe laser is set to $f_{\mathrm{rep}}=160\,\mathrm{MHz}$. Spin dynamics at frequencies significantly higher than the detector bandwidth are  measured without any loss of sensitivity. The investigated system is $n$-type bulk GaAs at the metal-to-insulator transition ($n_{\mathrm{D}}=1.8\times 10^{16}\,\mathrm{cm^{-3}}$). A crossover from homogeneous to inhomogeneous spin dephasing, i.e., from a Lorentzian to a Gaussian line shape, occurs at high magnetic fields (see Sec.~\ref{spindyn:investigation}). Spectra are shifted for clarity. The negative spin noise peak results from background noise subtraction. }
   \label{fig:ghz}
\end{figure}
SNS utilizing continuous-wave  lasers as in Fig.~\ref{fig:fig4-setup} can only  measure spin noise at frequencies below the  detector bandwidth and has so far been only been demonstrated  at frequencies smaller than  1~GHz.
Recently,  this limitation has been overcome by replacing the continuous-wave laser in Fig.~\ref{fig:fig4-setup}  with an ultrafast  pulsed laser light source \cite{mueller:prb:81:121202:2010}. Thereby, the spin-spin correlation in Eq.~(\ref{eq:correlation}) is only probed when an ultrashort laser pulse traverses the sample and the relevant correlator  additionally contains the probing pulse train:  
\begin{equation}
\langle s_z(0)s_z(t)\rangle\rightarrow\langle
s_z(0)s_z(t)\rangle\times \sum_n\delta\left( t-n/
f_{\mathrm{rep}}\right),
\end{equation}
where $f_{\mathrm{rep}}$ is the repetition rate of the laser source. Thus, the spin noise spectrum, which is given by a peak $S(f)$ around the Larmor frequency $\omega_{\mathrm{L}}/2\pi$ in conventional SNS, evolves into a sum of peaks all shifted by the repetition rate of
the laser:
\begin{equation}
S(f)\rightarrow\sum_{\pm m}S\left(f-m f_{\mathrm{rep}}\right).
\end{equation}
Accordingly, spin noise at frequencies much higher than the bandwidth of the detector appears to slow down due to this stroboscopic sampling and can still be detected. This new experimental technique of GHz SNS is applied in Ref.~\cite{mueller:prb:81:121202:2010} to detect spin noise at Larmor frequencies up to 16~GHz (see Fig.~\ref{fig:quality}). GHz SNS is limited to dynamics on timescales that are long with respect to the pulse length. Thus, sub ps pulses allow to access the THz regime. It is important to note, that this ultrafast sampling does not \textit{per se} introduce any further noise and, correspondingly, does not show a reduced sensitivity compared to conventional SNS. From a technical point of view, pulsed laser light sources generally tend  to a higher degree of instability than continuous-wave lasers; nevertheless, for the here discussed experiment, the resulting classical noise occurs on the frequency scale well below 1~Hz and is, hence, irrelevant to the experimental sensitivity. The maximal spin dephasing rates that can be resolved by this technique are limited by half of the laser repetition rate as well as the bandwidth of the detector.

 Starosielec and H\"agele suggested   ultrafast SNS, also  employing pulsed laser light \cite{starosielec:apl:93:051116:2008}. In their proposal, the spin-spin correlation function is  not investigated by means of frequency analysis, but in a more direct fashion by varying the time delay between two subsequent probe pulses . Experimental realization of this proposal would allow to detect spin dynamics with precessional frequencies and dephasing rates both only limited by the inverse pulse length.

\section{Applications}\label{applications}
The original motivation to transfer SNS to semiconductors was to implement a perturbation-free experimental probe to gain a better understanding of semiconductor spin dynamics that may help to realize spintronic devices. Besides from that, a new experimental method often carries some potential in itself to find its way from the laboratory towards applications. The technique of nuclear magnetic resonance is of course a great example for such a transfer and shows that it is in any case worthwhile to think about the potential of SNS.
In this section, two  potential applications of semiconductor SNS are reviewed. In the first application, SNS is employed as a quantum random number generator (Sec.~\ref{applications:qrng}). Secondly,  the spatial resolution of SNS can be utilized for sample characterization by acquiring three-dimensional images of the doping concentration (Sec.~\ref{applications:spatial}).

\subsection{Quantum Random Number Generator}\label{applications:qrng}
Pseudorandom numbers that are generated in deterministic computer algorithms may lead to erroneous results in numerical simulations \cite{ferrenberg:prl:69:3382:1992}. This problem can be circumvented by application of physical random number generators. Of course, actual randomness can only be achieved if the number generator relies on a truly unpredictable physical process. Quantum measurements are known to be inherently unpredictable and, hence, produce real random numbers. Katsoprinakis \textit{et al.} implemented a   quantum random number generator  based on spin noise measurements of Rubidium vapor where the bit rate of generated random numbers is given by the spin dephasing rate \cite{katsoprinakis:pra:77:054101:2008}. Hence, they argue that a quantum random  number generator based on semiconductor SNS may produce relatively high bit rates on the order of 10 Mbit/s.
\subsection{Spatially Resolved Measurements}\label{applications:spatial}
\begin{figure}[tb!]
   \centering
       \includegraphics[width=1.00\columnwidth]{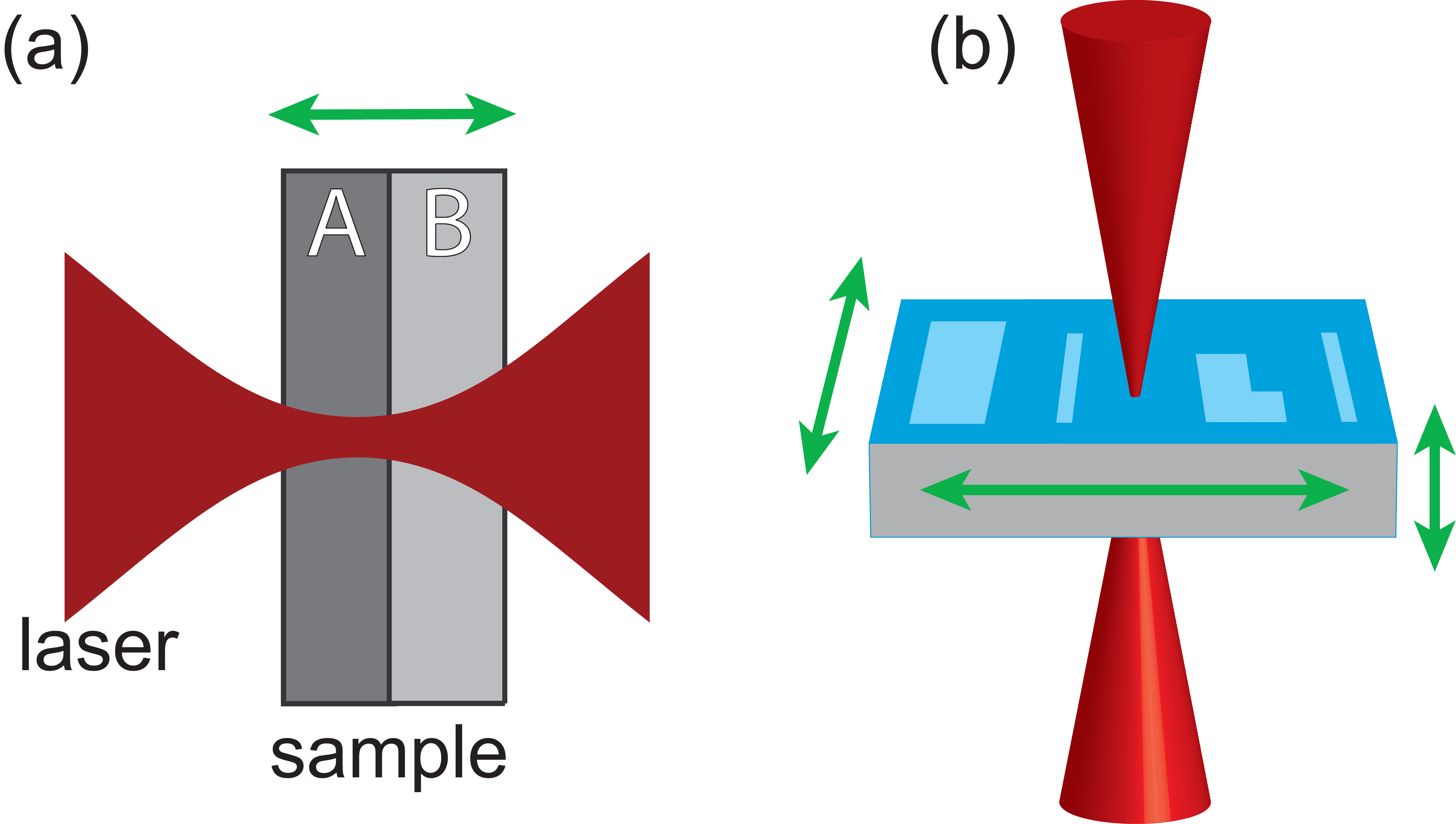}
   \caption{(a) Proof-of-principle experiment by R\"omer \textit{et al.} \cite{roemer:apl:94:112105:2009}: The thicknesses of the two different wafers A and B are measured by depth-resolved SNS. (b) SNS allows to produce three dimensional images of the doping concentration in a given semiconductor sample.}
   \label{fig:application}
\end{figure}
The spatial distribution of impurity atoms crucially determines the functional capability of semiconductor devices. With decreasing device size, even the stochastic dopant fluctuations can become relevant. However, the most often used  method to determine dopant concentrations are Hall measurements that have almost no spatial resolution. Secondary ion mass spectroscopy allows to map the impurity distribution, but is destructive. Scanning tunneling microscopy facilitates non-destructive investigations of the impurity distribution with atomic resolution, though, it is limited to the sample surface.  Now, SNS promises to close the gap between those methods that lack three-dimensional resolution  and those that are destructive.

SNS is not only sensitive to the spin dynamics at the sample surface as other optical techniques since SNS employs  below band gap light.  Furthermore, most of the spin noise signal is acquired within the Rayleigh range of the focused probe laser light. These two facts allow  to spatially resolve semiconductor spin dynamics in all three dimensions of space via SNS with strongly focused probe light.
In GaAs, the effective $g$-factor (see Ref. \cite{yang:prb:47:6807:1993} and references therein), the spin dephasing time (see Sec. \ref{spindyn:investigation}), as well as the spin noise power (see Secs.~\ref{sns:loc} and \ref{sns:deloc}) depend on the local doping concentration. These quantities specify the detected spin noise spectra and, therefore, spatially resolved SNS should allow to produce three dimensional images of the impurity concentration in a semiconductor sample (see Fig.~\ref{fig:application}~(b)). In 2009, this feature was demonstrated  in a proof-of-principle experiment \cite{roemer:apl:94:112105:2009}:  R\"omer and co-workers acquired a series of  spin noise spectra in a sample stack consisting of two different commercial $n$-doped GaAs wafers. The probe light was focused by  high aperture optics and the sample was axially scanned by varying the  focus position  (see Fig.~\ref{fig:application}~(a)). The contributions of the two individual samples can be recovered from the spin noise spectra by a fitting routine. That way, the thickness of the two wafers can be correctly reproduced, i.e., a spatial doping profile is reconstructed. Even a three-dimensional mapping of the doping concentration can, in principle, be achieved by simultaneous laterally and depth-resolved measurements. The spatial resolution may be extended well beyond the limits of the laser focus and the Rayleigh range if the laser spot scans the sample by small steps in conjunction with sophisticated  data processing.


\section{Outlook}\label{outlook}


SNS allows in principle perturbation-free investigation of spin dynamics in semiconductors and is in this regard a unique  experimental tool.  Application of SNS is primarily useful for sample systems in which excitations strongly change the investigated dynamics, as in low doped semiconductors at low temperatures and in systems where all well-known spin dephasing processes are known to be inefficient.  So far SNS has been applied to $n$-type bulk GaAs \cite{romer:rsi:78:103903:2007, roemer:apl:94:112105:2009, crooker:prb:79:035208:2009, romer:prb:81:075216:2010, mueller:prb:81:121202:2010}, GaAs/AlGaAs based quantum wells \cite{muller:prl:101:206601:2008}, and ensembles of (In,Ga)As/GaAs quantum dots \cite{crooker:prl:104:036601:2010}. Nevertheless, SNS can be universally utilized in   other semiconductor materials---with  direct as well as with  indirect optical transitions---and  should also work in other solid state material classes, e.g, in materials with magnetic order where collective magnetic modes are thermally excited.

SNS probes the spin fluctuations in the investigated sample system, i.e., the spin-spin correlation function. Spin correlations  of higher order, which are also contained in the acquired time signal, can reveal further information about the underlying spin dynamics. For example, third-order correlations may allow for separation of homogeneous spin dephasing from inhomogeneous processes  in prospective SNS experiments on very small electron ensembles \cite{liu:njp:12:013018:2010}.

The experimental sensitivity of semiconductor SNS is in the case of large electron ensembles, where moderately high probe laser powers can be utilized,  mostly limited by optical shot noise. Here, application of squeezed light as probe light can further increase  the signal-to-noise ratio \cite{sorenson:prl:80:3487:1998}. In the case of small electron ensembles, significant noise contributions from the detector cannot be avoided and  further enhancement of the experimental sensitivity  may be achieved by  a Mach-Zehnder interferometer like setup \cite{appel:pnas:106:10960:2009,sugita2003,laforge:apl:91:121115:2007, laforge:rsi:79:063106:2008} or cavity enhancement of the Faraday rotation \cite{salis:prb:72:115325:2005, li:apl:88:193126:2006}.

The role of residual sample excitations, e.g., considering optical excitation of deep centers \cite{mueller:prb:81:121202:2010} or spin flip light scattering \cite{gorbovitskii:optspec:54:229:1983,atature:natphys:3:101:2007}, clearly needs further attention. This question is of particular interest with respect to a possible transfer of the quantum non-demolition experiments on atomic gases (see Sec. \ref{sns:atomic})  to semiconductor physics  as well as for the detection of  a single electronic spin confined in a quantum dot. Only if spin flip scattering of the probe light occurs on timescales longer than the spin lifetime, SNS can fulfill  a meaningful   measurement of a single electronic spin. The realization of such a quantum non-demolition measurement in a semiconductor has recently gained a lot of research interest (see, e.g., Refs.~\cite{sarovar:prb:78:245302:2008,atature:natphys:3:101:2007,bulaevskii:prl:92:177001:2004,  kim:prl:101:236804:2008,  wabnig:njp:11:043031:2009}). Especially, an implementation based on optical detection via Faraday rotation---as in SNS---is desirable since  such a measurement is employed as building block in several schemes for photon-spin and spin-spin entanglement (see, e.g., Refs.~ \cite{leuenberger:prl:94:107401:2005, leuenberger:prb:73:075312:2006, grond:prb:77:165307:2008, hu:prb:78:085307:2008,potz:prb:77:035310:2008, seigneur:jap:104:014307:2008}).
